\documentclass[preprint,aps,prd,USpaper,nofootinbib,showpacs,superscriptaddress]{revtex4-2}
\usepackage{amsmath,amssymb,graphicx,mathrsfs,xcolor}
\usepackage{bm} 
\usepackage{feynmp}

\usepackage{ifpdf}
\ifpdf
\fi
\makeatletter
\def\endfmffile{%
  \fmfcmd{\p@rcent\space the end.^^J%
          end.^^J%
          endinput;}%
  \if@fmfio
    \immediate\closeout\@outfmf
  \fi
  \IfFileExists{\thefmffile.mp}{\immediate\write18{mpost \thefmffile}}{}
  \let\thefmffile\relax
}
\makeatother
\newcommand{\nn}{\nonumber\\}

\newcommand{\Vg}{V_{\mbox{\scriptsize gluon}} }
\newcommand{\Vc}{V_{\mbox{\scriptsize conf}}}

\newcommand{\bGamma}{\bar{\Gamma}}

\newcommand{\ben}{\begin{displaymath}}
\newcommand{\een}{\end{displaymath}}
\newcommand{\be}{\begin{equation}}
\newcommand{\ee}{\end{equation}}
\newcommand{\bea}{\begin{eqnarray}}
\newcommand{\eea}{\end{eqnarray}}

\newcommand{\A}{\alpha}

\newcommand{\Gf}{G}

\newcommand{\q}{\bar{q}}
\newcommand{\D}{\bar{D}}

\newcommand{\PPhi}{{\it\Phi}}
\newcommand{\GG}{{\it\Gamma}}
\newcommand{\bGG}{{\it \bar\Gamma}}
\newcommand{\RR}{{\cal R}}
\newcommand{\PP}{{\cal P}}
\newcommand{\TT}{{\cal T}}

\newcommand{\bc}{\begin{center}}
\newcommand{\ec}{\end{center}}

\newcommand{\eqn}[1]{\label{#1}}
\newcommand{\eq}[1]{Eq.~(\ref{#1})}
\newcommand{\eqs}[1]{Eqs.~(\ref{#1})}
\newcommand{\fign}[1]{\label{#1}}
\newcommand{\fig}[1]{Fig.~\ref{#1}}

\begin{document}
\title{Exact unified tetraquark equations}
\author{B. Blankleider}
\email{boris.blankleider@flinders.edu.au}
\affiliation{
College of Science and Engineering,
 Flinders University, Bedford Park, SA 5042, Australia}
\author{A. N. Kvinikhidze}
\email{sasha\_kvinikhidze@hotmail.com}
\affiliation{
College of Science and Engineering,
 Flinders University, Bedford Park, SA 5042, Australia}
\affiliation{Andrea Razmadze Mathematical Institute of Tbilisi State University, 6, Tamarashvili Str., 0186 Tbilisi, Georgia}

\date{\today}

\begin{abstract}
Recently we formulated covariant equations describing the tetraquark in terms of an admixture of two-body states $D\bar D$ (diquark-antidiquark), $MM$ (meson-meson), and three-body-like states where two of the quarks are spectators while the other two are interacting [Phys.\ Rev.\ D 107, 094014 (2023)]. A feature of these equations is that they unify descriptions of seemingly unrelated models of the tetraquark, like, for example, the $D\bar D$ model of the Moscow group [Faustov {\it et al.}, Universe {\bf 7}, 94 (2021)] and the coupled channel $D \bar D-MM$ model of the Giessen group [Heupel {\it et al.}, Phys.\ Lett.\ {\bf B718}, 545 (2012)]. Here we extend these equations to the exact case where $q\q$ annihilation is incorporated explicitly, and all previously neglected terms (three-body forces, non-pole contributions to two-quark t matrices, etc.) are taken into account through the inclusion of a single $q\q$ potential $\Delta$.  
\end{abstract}

\maketitle
\newpage

\section{Introduction}

In 1992, Khvedelidze and Kvinikhidze (KK) formulated exact covariant equations for four particles interacting via pairwise interactions \cite{Khvedelidze:1991qb}.  A feature of these equations is that pairwise interactions enter the theory not only via single  two-body kernels $K_a$ describing the scattering of a particle pair $a$ while particle pair $a'$ is spectating, but also via subtractions of products of kernels $K_a K_{a'}$, which are needed to avoid overcounting, see \eq{Kaa} below. However, the equations are rearranged in such a way 
 that all two-body kernels, including these subtraction terms, disappear, with two-body interactions entering the full theory as a {\it sum} of single  two-body t matrices $T_a$ and  products of t matrices $T_aT_{a'}$. In particular, the final equations use as input the amplitudes $T_{aa'}$, defined as the full 4-body t matrix where all interactions are switched off except those within the pairs labelled by $a$ and $a'$, and which themselves are given in terms of $T_a$ and $T_{a'}$ as
\be
T_{aa'} = T_a G_{a'}^{0}{}^{-1}+T_{a'}G_{a}^{0}{}^{-1}  + T_a T_{a'}  \eqn{Taa}
\ee
where $G_{a}^0$ and $G_{a'}^0$ are the Green functions describing the free propagation of particle-pairs $a$ and $a'$, respectively \cite{Khvedelidze:1991qb,Heupel:2012ua}. 

In 2012, members of the Giessen group, Heupel {\it et al.} \cite{Heupel:2012ua}, applied KK's formulation to the case of two quarks ($2q$) and two antiquarks ($2\q$), introducing the further simplifications where the first two terms on the right hand side (rhs) of \eq{Taa} are neglected, while the two-body t matrices in the last term $T_a T_{a'}$ are approximated by their meson ($M$), diquark ($D$), and antidiquark ($\D$) pole contributions. The resulting equations were then reduced to a set of coupled two-body equations for the $MM$-tetraquark and $D\bar D$-tetraquark amplitudes $\phi_M$ and $\phi_D$, respectively, as illustrated in \fig{GBSE}. A feature of the Giessen group's approach is that it is based on a rigorous field-theoretic derivation for the $2q2\bar q$ system where all approximations are clearly specified. Numerical solutions of these equations have been pursued by this group in a series of recent publications  \cite{Heupel:2012ua,Eichmann:2015cra,Eichmann:2020oqt, Santowsky:2021bhy}.
\begin{figure}[t]
\begin{center}
\begin{fmffile}{no2q}
\begin{align*}
\parbox{10mm}{
\begin{fmfgraph*}(10,13)
\fmfset{dash_len}{2mm}
\fmfstraight
\fmfleftn{l}{7}\fmfrightn{r}{7}\fmfbottomn{b}{6}\fmftopn{t}{6}
\fmf{phantom}{l2,mb2,r2}
\fmf{phantom}{l6,mt2,r6}
\fmffreeze
\fmf{phantom}{r2,phi,r6}
\fmffreeze
\fmf{dashes}{phi,l2}
\fmf{dashes}{phi,l6}
\fmfv{d.sh=circle,d.f=empty,d.si=22,label=$\hspace{-4.5mm}\phi_M$,background=(1,,.51,,.5)}{phi}
\end{fmfgraph*}}
\hspace{6mm} &= \hspace{2mm}
\parbox{28mm}{
\begin{fmfgraph*}(28,13)
\fmfset{dash_len}{2.2mm}
\fmfstraight
\fmfleftn{l}{7}\fmfrightn{r}{7}\fmfbottomn{b}{6}\fmftopn{t}{6}
\fmf{phantom}{l2,vb1,mb1,vb2,mb2,r2}
\fmf{phantom}{l6,vt1,mt1,vt2,mt2,r6}
\fmf{phantom}{l2,xb1,mb1,xb2,mb2,r2}
\fmf{phantom}{l6,xt1,mt1,xt2,mt2,r6}
\fmf{phantom}{l2,yb1,mb1,yb2,mb2,r2}
\fmf{phantom}{l6,yt1,mt1,yt2,mt2,r6}
\fmffreeze
\fmf{phantom}{r2,phi,r6}
\fmffreeze
\fmfshift{3 right}{xb1}
\fmfshift{3 down}{xb1}
\fmfshift{3 left}{xt2}
\fmfshift{3 down}{xt2}
\fmfshift{3 right}{xt1}
\fmfshift{3 down}{xt1}
\fmfshift{3 left}{xb2}
\fmfshift{3 down}{xb2}
\fmfshift{3 right}{yb1}
\fmfshift{3 up}{yb1}
\fmfshift{3 left}{yt2}
\fmfshift{3 up}{yt2}
\fmfshift{3 right}{yt1}
\fmfshift{3 up}{yt1}
\fmfshift{3 left}{yb2}
\fmfshift{3 up}{yb2}
\fmfv{d.sh=circle,d.f=empty,d.si=8,background=(1,,.51,,.5)}{vb1}
\fmfv{d.sh=circle,d.f=empty,d.si=8,background=(1,,.51,,.5)}{vb2}
\fmfv{d.sh=circle,d.f=empty,d.si=8,background=(1,,.51,,.5)}{vt1}
\fmfv{d.sh=circle,d.f=empty,d.si=8,background=(1,,.51,,.5)}{vt2}
\fmf{dashes}{l2,vb1}
\fmf{dashes}{l6,vt1}
\fmf{dashes}{phi,xb2}
\fmf{dashes}{phi,yt2}
\fmf{phantom}{vb1,vb2}
\fmf{phantom}{vt1,vt2}
\fmfset{arrow_len}{2.2mm}
\fmf{plain,rubout=3}{yb1,xt2}
\fmf{plain}{xt1,yb2}
\fmf{phantom_arrow}{xt1,c}
\fmf{phantom_arrow}{c,yb2}
\fmf{phantom_arrow,rubout=2}{yb1,c}
\fmf{phantom_arrow,rubout=2}{c,xt2}
\fmf{fermion}{yt2,yt1}
\fmf{fermion}{xb2,xb1}
\fmfv{d.sh=circle,d.f=empty,d.si=22,label=$\hspace{-4.5mm}\phi_M$,background=(1,,.51,,.5)}{phi}
\end{fmfgraph*}}
\hspace{7mm}+\hspace{1mm}
\parbox{28mm}{
\begin{fmfgraph*}(28,13)
\fmfstraight
\fmfleftn{l}{7}\fmfrightn{r}{7}\fmfbottomn{b}{6}\fmftopn{t}{6}
\fmf{phantom}{l2,vb1,mb1,vb2,mb2,r2}
\fmf{phantom}{l6,vt1,mt1,vt2,mt2,r6}
\fmf{phantom}{l2,xb1,mb1,xb2,mb2,r2}
\fmf{phantom}{l6,xt1,mt1,xt2,mt2,r6}
\fmf{phantom}{l2,yb1,mb1,yb2,mb2,r2}
\fmf{phantom}{l6,yt1,mt1,yt2,mt2,r6}
\fmffreeze
\fmf{phantom}{r2,phi,r6}
\fmffreeze
\fmfshift{3 right}{xb1}
\fmfshift{3 down}{xb1}
\fmfshift{3 left}{xt2}
\fmfshift{3 down}{xt2}
\fmfshift{3 right}{xt1}
\fmfshift{3 down}{xt1}
\fmfshift{3 left}{xb2}
\fmfshift{3 down}{xb2}
\fmfshift{3 right}{yb1}
\fmfshift{3 up}{yb1}
\fmfshift{3 left}{yt2}
\fmfshift{3 up}{yt2}
\fmfshift{3 right}{yt1}
\fmfshift{3 up}{yt1}
\fmfshift{3 left}{yb2}
\fmfshift{3 up}{yb2}
\fmfv{d.sh=circle,d.f=empty,d.si=8,background=(1,,.51,,.5)}{vb1}
\fmfv{d.sh=circle,d.f=empty,d.si=8,background=(.6235,,.7412,,1)}{vb2}
\fmfv{d.sh=circle,d.f=empty,d.si=8,background=(1,,.51,,.5)}{vt1}
\fmfv{d.sh=circle,d.f=empty,d.si=8,background=(.6235,,.7412,,1)}{vt2}
\fmf{dashes}{l2,vb1}
\fmf{dashes}{l6,vt1}
\fmf{dbl_plain}{phi,xb2}
\fmf{dbl_plain}{phi,yt2}
\fmf{phantom}{vb1,vb2}
\fmf{phantom}{vt1,vt2}
\fmfset{arrow_len}{2.2mm}
\fmf{plain,rubout=3}{xt2,yb1}
\fmf{plain}{yb2,xt1}
\fmf{phantom_arrow,rubout=2}{xt2,c}
\fmf{phantom_arrow,rubout=2}{c,yb1}
\fmf{phantom_arrow}{xt1,c}
\fmf{phantom_arrow}{c,yb2}
\fmf{fermion}{yt2,yt1}
\fmf{fermion}{xb1,xb2}
\fmfv{d.sh=circle,d.f=empty,d.si=22,label=$\hspace{-4.5mm}\phi_D$,background=(.6235,,.7412,,1)}{phi}
\end{fmfgraph*}}\\[6mm]
\parbox{10mm}{
\begin{fmfgraph*}(10,13)
\fmfstraight
\fmfleftn{l}{7}\fmfrightn{r}{7}\fmfbottomn{b}{6}\fmftopn{t}{6}
\fmf{phantom}{l2,mb2,r2}
\fmf{phantom}{l6,mt2,r6}
\fmffreeze
\fmf{phantom}{r2,phi,r6}
\fmffreeze
\fmf{dbl_plain}{phi,l2}
\fmf{dbl_plain}{phi,l6}
\fmfv{d.sh=circle,d.f=empty,d.si=22,label=$\hspace{-4.5mm}\phi_D$,background=(.6235,,.7412,,1)}{phi}
\end{fmfgraph*}}
\hspace{6mm} &= \hspace{2mm}
\parbox{28mm}{
\begin{fmfgraph*}(28,13)
\fmfset{dash_len}{2.2mm}
\fmfstraight
\fmfleftn{l}{7}\fmfrightn{r}{7}\fmfbottomn{b}{6}\fmftopn{t}{6}
\fmf{phantom}{l2,vb1,mb1,vb2,mb2,r2}
\fmf{phantom}{l6,vt1,mt1,vt2,mt2,r6}
\fmf{phantom}{l2,xb1,mb1,xb2,mb2,r2}
\fmf{phantom}{l6,xt1,mt1,xt2,mt2,r6}
\fmf{phantom}{l2,yb1,mb1,yb2,mb2,r2}
\fmf{phantom}{l6,yt1,mt1,yt2,mt2,r6}
\fmffreeze
\fmf{phantom}{r2,phi,r6}
\fmffreeze
\fmfshift{3 right}{xb1}
\fmfshift{3 down}{xb1}
\fmfshift{3 left}{xt2}
\fmfshift{3 down}{xt2}
\fmfshift{3 right}{xt1}
\fmfshift{3 down}{xt1}
\fmfshift{3 left}{xb2}
\fmfshift{3 down}{xb2}
\fmfshift{3 right}{yb1}
\fmfshift{3 up}{yb1}
\fmfshift{3 left}{yt2}
\fmfshift{3 up}{yt2}
\fmfshift{3 right}{yt1}
\fmfshift{3 up}{yt1}
\fmfshift{3 left}{yb2}
\fmfshift{3 up}{yb2}
\fmfv{d.sh=circle,d.f=empty,d.si=8,background=(.6235,,.7412,,1)}{vb1}
\fmfv{d.sh=circle,d.f=empty,d.si=8,background=(1,,.51,,.5)}{vb2}
\fmfv{d.sh=circle,d.f=empty,d.si=8,background=(.6235,,.7412,,1)}{vt1}
\fmfv{d.sh=circle,d.f=empty,d.si=8,background=(1,,.51,,.5)}{vt2}
\fmf{dbl_plain}{l2,vb1}
\fmf{dbl_plain}{l6,vt1}
\fmf{dashes}{phi,xb2}
\fmf{dashes}{phi,yt2}
\fmf{phantom}{vb1,vb2}
\fmf{phantom}{vt1,vt2}
\fmfset{arrow_len}{2.2mm}
\fmf{plain,rubout=3}{yb1,xt2}
\fmf{plain}{yb2,xt1}
\fmf{phantom_arrow,rubout=2}{yb1,c}
\fmf{phantom_arrow,rubout=2}{c,xt2}
\fmf{phantom_arrow}{yb2,c}
\fmf{phantom_arrow}{c,xt1}
\fmf{fermion}{yt2,yt1}
\fmf{fermion}{xb1,xb2}
\fmfv{d.sh=circle,d.f=empty,d.si=22,label=$\hspace{-4.5mm}\phi_M$,background=(1,,.51,,.5)}{phi}
\end{fmfgraph*}}
\end{align*}
\end{fmffile}   
\vspace{-3mm}

\caption{\fign{GBSE}  Tetraquark equations of the Giessen group \cite{Heupel:2012ua,Eichmann:2015cra,Eichmann:2020oqt, Santowsky:2021bhy}. Form factor $\phi_M$ couples the tetraquark to two mesons (dashed lines), and form factors $\phi_D$ couples the tetraquark to diquark-antidiquark states (double-lines). Quarks (antiquarks) are represented by left (right) directed lines.}
\end{center}
\end{figure}
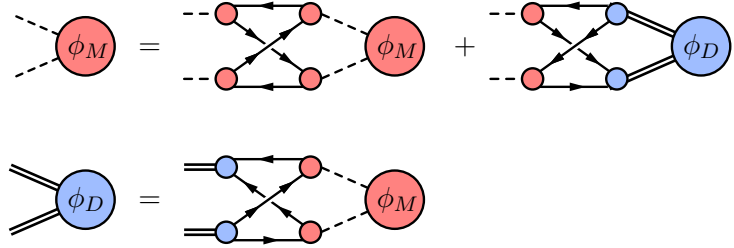

A separate approach has been followed for many years by the Moscow group, Faustov {\it et al.}\! \cite{Ebert:2005nc,Faustov:2020qfm,Faustov:2021hjs,Faustov:2022mvs}, who model tetraquarks as a diquark-antidiquark  system. The Moscow group's model can be viewed as being based on the solutions of the bound-state equation for the $D\bar D$-tetraquark amplitude $\phi_D$, as illustrated in \fig{MBSE}. As seen from this figure, the kernel of the equation consists of a single term where a $q\q$ pair scatters elastically in the presence of spectating $q$ and $\q$ quarks. More specifically, the Moscow model corresponds to the case where $T_{q\q}$, the t matrix describing the mentioned $q\q$ scattering, is expressed as a sum of two potentials
\be
T_{q\q} = V_{\mbox{\scriptsize gluon}} + V_{\mbox{\scriptsize conf}}   \eqn{t_moscow}
\ee
where $\Vg$ is the $q\q$ one-gluon-exchange potential and $\Vc$ is a local confining  potential.\footnote{ 
To be precise, the Moscow group uses quasipotential bound state form factors instead of the $D\rightarrow qq$ form factor $\Gamma_{12}(p,P)$ and the $\bar D \rightarrow \q\q$ form factor $\Gamma_{34}(p,P)$, appearing as small blue circles in \fig{MBSE}. Formally, this is equivalent to assuming that $\Gamma_{12}(p,P)$ and $\Gamma_{34}(p,P)$  do not depend on the longitudinal projection of the relative 4-momentum $p$ with respect to the total momentum $P$ of the two quarks or two antiquarks.}

\begin{figure}[b]
\begin{center}
\begin{fmffile}{faust}
\[
\parbox{10mm}{
\begin{fmfgraph*}(10,13)
\fmfstraight
\fmfleftn{l}{7}\fmfrightn{r}{7}\fmfbottomn{b}{6}\fmftopn{t}{6}
\fmf{phantom}{l2,mb2,r2}
\fmf{phantom}{l6,mt2,r6}
\fmffreeze
\fmf{phantom}{r2,phi,r6}
\fmffreeze
\fmf{dbl_plain}{phi,l2}
\fmf{dbl_plain}{phi,l6}
\fmfv{d.sh=circle,d.f=empty,d.si=22,label=$\hspace{-4.5mm}\phi_D$,background=(.6235,,.7412,,1)}{phi}
\end{fmfgraph*}}
\hspace{6mm} = \hspace{5mm}
\parbox{30mm}{
\begin{fmfgraph*}(30,13)
\fmfstraight
\fmfleftn{l}{7}\fmfrightn{r}{7}\fmfbottomn{b}{7}\fmftopn{t}{7}
\fmf{phantom}{l2,vb1,mb1,vb2,mb2,r2}
\fmf{phantom}{l6,vt1,mt1,vt2,mt2,r6}
\fmf{phantom}{l2,xb1,mb1,xb2,mb2,r2}
\fmf{phantom}{l6,xt1,mt1,xt2,mt2,r6}
\fmf{phantom}{l2,yb1,mb1,yb2,mb2,r2}
\fmf{phantom}{l6,yt1,mt1,yt2,mt2,r6}
\fmf{phantom}{l4,l4a,l4b,r4b,r4a,r4}
\fmffreeze
\fmf{phantom}{l4,c,r4a}
\fmf{phantom}{l4,clb,r4a}
\fmf{phantom}{l4,clt,r4a}
\fmf{phantom}{l4,crb,r4a}
\fmf{phantom}{l4,crt,r4a}
\fmffreeze
\fmf{phantom}{r2,phi,r6}
\fmffreeze
\fmfshift{8 left}{phi}
\fmfshift{3 right}{xb1}
\fmfshift{3 down}{xb1}
\fmfshift{3 left}{xt2}
\fmfshift{3 down}{xt2}
\fmfshift{3 right}{xt1}
\fmfshift{3 down}{xt1}
\fmfshift{3 left}{xb2}
\fmfshift{3 down}{xb2}
\fmfshift{3 right}{yb1}
\fmfshift{3 up}{yb1}
\fmfshift{3 left}{yt2}
\fmfshift{3 up}{yt2}
\fmfshift{3 right}{yt1}
\fmfshift{3 up}{yt1}
\fmfshift{3 left}{yb2}
\fmfshift{3 up}{yb2}
\fmfshift{3 left}{clb}
\fmfshift{3 down}{clb}
\fmfshift{3 left}{clt}
\fmfshift{3 up}{clt}
\fmfshift{3 right}{crb}
\fmfshift{3 down}{crb}
\fmfshift{3 right}{crt}
\fmfshift{3 up}{crt}
\fmfv{d.sh=circle,d.f=empty,d.si=8,background=(.6235,,.7412,,1)}{vb1}
\fmfv{d.sh=circle,d.f=empty,d.si=8,background=(.6235,,.7412,,1)}{vb2}
\fmfv{d.sh=circle,d.f=empty,d.si=8,background=(.6235,,.7412,,1)}{vt1}
\fmfv{d.sh=circle,d.f=empty,d.si=8,background=(.6235,,.7412,,1)}{vt2}
\fmfv{d.sh=circle,d.f=empty,d.si=8,background=(1,,.51,,.5)}{c}
\fmf{dbl_plain}{l2,vb1}
\fmf{dbl_plain}{l6,vt1}
\fmf{dbl_plain}{phi,xb2}
\fmf{dbl_plain}{phi,yt2}
\fmfset{arrow_len}{2.2mm}
\fmfi{fermion}{vloc(__xt2) .. vloc(__crt)}
\fmfi{fermion}{ vloc(__clt)..vloc(__xt1)}
\fmfi{fermion}{vloc(__crb) ..vloc(__yb2)}
\fmfi{fermion}{vloc(__yb1) ..vloc(__clb)}
\fmfi{fermion}{vloc(__yt2) .. vloc(__yt1)}
\fmfi{fermion}{vloc(__xb1) .. vloc(__xb2)}
\fmfv{d.sh=circle,d.f=empty,d.si=22,label=$\hspace{-4.5mm}\phi_D$,background=(.6235,,.7412,,1)}{phi}
\end{fmfgraph*}}
\]
\end{fmffile}   
\vspace{-3mm}

\caption{\fign{MBSE}  Diquark-antidiquark bound state equation encompassing the Moscow group's approach \cite{Ebert:2005nc,Faustov:2020qfm,Faustov:2021hjs,Faustov:2022mvs}.  The form factor $\phi_D$ couples the tetraquark to diquark and antidiquark states (both represented by double-lines). Shown is the general form of the kernel where one $q \bar q$ pair interacts (the red circle representing the corresponding t matrix $T_{q\q}$) while the other $q \bar q$ pair is spectating. }
\end{center}
\end{figure}
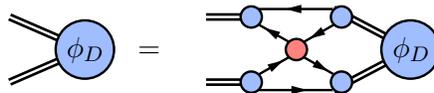

Recently we have shown how these two seemingly unrelated approaches to the tetraquark can be unified by providing a common theoretical basis for the two separate models \cite{Kvinikhidze:2023djp}. The key idea is based on the observation that the first two terms of \eq{Taa}, if retained in the application of KK's formalism to the $2q2\q$ system, would generate three-body-like intermediate states of the form $q q (T_{\q\q})$, $\q\q (T_{qq})$, and $q\bar q (T_{q\q})$, where two of the quarks are spectators while the other two are interacting, the last of which, namely $q\q (T_{q\q})$, would account for the kernel of the Moscow group's bound state equation. Thus, in Ref.\ \cite{Kvinikhidze:2023djp}, we applied KK's equations to the $2q2\q$ system, similarly to  Heupel {\it et al.} \cite{Heupel:2012ua}, but we retained the first two terms on the rhs of \eq{Taa}.  In this way we obtained equations for the tetraqaurk  which can be expressed in matrix form as
\be
\phi = \left[V^{(0)} + V^{(1)}+\dots \right] D \phi    \eqn{eq}
\ee
where $\phi$ is a column of the tetraquark form factors, $D$ is a diagonal matrix consisting of the two-meson propagator $MM$ and the diquark-antidiquark propagator $D\D$, namely,
\be
\phi = \begin{pmatrix} \phi_M \\  \phi_D \end{pmatrix}
=
\begin{fmffile}{phi}
\begin{pmatrix} 
\parbox{10mm}{
\begin{fmfgraph*}(10,13)
\fmfset{dash_len}{2mm}
\fmfstraight
\fmfleftn{l}{7}\fmfrightn{r}{7}\fmfbottomn{b}{6}\fmftopn{t}{6}
\fmf{phantom}{l2,mb2,r2}
\fmf{phantom}{l6,mt2,r6}
\fmffreeze
\fmf{phantom}{r2,phi,r6}
\fmffreeze
\fmf{dashes}{phi,l2}
\fmf{dashes}{phi,l6}
\fmfv{d.sh=circle,d.f=empty,d.si=22,label=$\hspace{-4.5mm}\phi_{M}$,background=(1,,.51,,.5)}{phi}
\end{fmfgraph*}} \hspace{4mm} \\
\parbox{10mm}{
\begin{fmfgraph*}(10,13)
\fmfstraight
\fmfleftn{l}{7}\fmfrightn{r}{7}\fmfbottomn{b}{6}\fmftopn{t}{6}
\fmf{phantom}{l2,mb2,r2}
\fmf{phantom}{l6,mt2,r6}
\fmffreeze
\fmf{phantom}{r2,phi,r6}
\fmffreeze
\fmf{dbl_plain}{phi,l2}
\fmf{dbl_plain}{phi,l6}
\fmfv{d.sh=circle,d.f=empty,d.si=22,label=$\hspace{-4.5mm}\phi_{D}$,background=(.6235,,.7412,,1)}{phi}
\end{fmfgraph*}}\hspace{3mm} 
\end{pmatrix}
\end{fmffile},
\hspace{1cm} D= \begin{pmatrix} \frac{1}{2} M M  & 0\\ 0 & D\D \end{pmatrix},    \eqn{phiD}
\ee
where we have used the same graphical representation of the tetraquark form factors as in Figs.\ 1 and 2, 
and where the kernel of the equation (term in square brackets) consists of an infinite series, the first two terms of which, $V^{(0)}$ and $V^{(1)}$, are defined in terms of Feynman graphs as shown in \fig{Feyn0} and \fig{Feyn1}, respectively. The ellipsis in \eq{eq} represents terms consisting of Feynman diagrams, similar to those of $V^{(1)}$, but describing the full multiple scattering of quarks in intermediate state. We refer to \eq{eq} as the ``unified tetraquark equations" \cite{Kvinikhidze:2023djp}.

That these equations unify the works of the Giessen and Moscow groups can be seen from the fact that the equations of Heupel {\it et al.} correspond to keeping just the term $V^{(0)}$ term in the kernel of \eq{eq}:
\be
\phi = V^{(0)} D \phi,           \hspace{1cm} \mbox{(Giessen group model)}     \eqn{phi_Giessen}
\ee
while the Moscow group model follows by keeping just the term $V^{(1)}$ term in the kernel of \eq{eq}:
\be
\phi = V^{(1)} D \phi,          \hspace{1cm} \mbox{(Moscow group model)}        \eqn{phi_Moscow}
\ee
but with all but the $D\D\rightarrow D\D$ elements of $V^{(1)}$ set to zero.
Our formulation suggests\ that the Giessen and Moscow groups have been investigating two non-overalpping aspects of the same model of the tetraquark. 

\begin{figure}[t]
\begin{center}
\begin{fmffile}{V0mat}
\begin{align*}
V^{(0)} \hspace{1mm} & = \hspace{2mm}
\begin{pmatrix}  \hspace{2mm}
\parbox{23mm}{
\begin{fmfgraph*}(23,13)
\fmfstraight
\fmfleftn{l}{7}\fmfrightn{r}{7}\fmfbottomn{b}{5}\fmftopn{t}{5}
\fmf{phantom}{l2,vb1,mb1,vb2,r2}
\fmf{phantom}{l6,vt1,mt1,vt2,r6}
\fmf{phantom}{l2,xb1,mb1,xb2,r2}
\fmf{phantom}{l6,xt1,mt1,xt2,r6}
\fmf{phantom}{l2,yb1,mb1,yb2,r2}
\fmf{phantom}{l6,yt1,mt1,yt2,r6}
\fmf{phantom}{t3,c,b3}
\fmffreeze
\fmfshift{3 right}{xb1}
\fmfshift{3 down}{xb1}
\fmfshift{3 left}{xt2}
\fmfshift{3 down}{xt2}
\fmfshift{3 right}{xt1}
\fmfshift{3 down}{xt1}
\fmfshift{3 left}{xb2}
\fmfshift{3 down}{xb2}
\fmfshift{3 right}{yb1}
\fmfshift{3 up}{yb1}
\fmfshift{3 left}{yt2}
\fmfshift{3 up}{yt2}
\fmfshift{3 right}{yt1}
\fmfshift{3 up}{yt1}
\fmfshift{3 left}{yb2}
\fmfshift{3 up}{yb2}
\fmfv{d.sh=circle,d.f=empty,d.si=8,background=(1,,.51,,.5)}{vb1}
\fmfv{d.sh=circle,d.f=empty,d.si=8,background=(1,,.51,,.5)}{vb2}
\fmfv{d.sh=circle,d.f=empty,d.si=8,background=(1,,.51,,.5)}{vt1}
\fmfv{d.sh=circle,d.f=empty,d.si=8,background=(1,,.51,,.5)}{vt2}
\fmf{dashes}{l2,vb1}
\fmf{dashes}{l6,vt1}
\fmf{dashes}{r2,vb2}
\fmf{dashes}{r6,vt2}
\fmfset{arrow_len}{2.2mm}
\fmf{plain,rubout=3}{yb1,xt2}
\fmf{plain}{xt1,yb2}
\fmf{phantom_arrow}{xt1,c}
\fmf{phantom_arrow}{c,yb2}
\fmf{phantom_arrow,rubout=2}{yb1,c}
\fmf{phantom_arrow,rubout=2}{c,xt2}
\fmf{fermion}{yt2,yt1}
\fmf{fermion}{xb2,xb1}
\end{fmfgraph*}}
&  \hspace{10mm}
\parbox{23mm}{
\begin{fmfgraph*}(23,13)
\fmfstraight
\fmfleftn{l}{7}\fmfrightn{r}{7}\fmfbottomn{b}{5}\fmftopn{t}{5}
\fmf{phantom}{l2,vb1,mb1,vb2,r2}
\fmf{phantom}{l6,vt1,mt1,vt2,r6}
\fmf{phantom}{l2,xb1,mb1,xb2,r2}
\fmf{phantom}{l6,xt1,mt1,xt2,r6}
\fmf{phantom}{l2,yb1,mb1,yb2,r2}
\fmf{phantom}{l6,yt1,mt1,yt2,r6}
\fmf{phantom}{t3,c,b3}
\fmffreeze
\fmfshift{3 right}{xb1}
\fmfshift{3 down}{xb1}
\fmfshift{3 left}{xt2}
\fmfshift{3 down}{xt2}
\fmfshift{3 right}{xt1}
\fmfshift{3 down}{xt1}
\fmfshift{3 left}{xb2}
\fmfshift{3 down}{xb2}
\fmfshift{3 right}{yb1}
\fmfshift{3 up}{yb1}
\fmfshift{3 left}{yt2}
\fmfshift{3 up}{yt2}
\fmfshift{3 right}{yt1}
\fmfshift{3 up}{yt1}
\fmfshift{3 left}{yb2}
\fmfshift{3 up}{yb2}
\fmfv{d.sh=circle,d.f=empty,d.si=8,background=(1,,.51,,.5)}{vb1}
\fmfv{d.sh=circle,d.f=empty,d.si=8,background=(.6235,,.7412,,1)}{vb2}
\fmfv{d.sh=circle,d.f=empty,d.si=8,background=(1,,.51,,.5)}{vt1}
\fmfv{d.sh=circle,d.f=empty,d.si=8,background=(.6235,,.7412,,1)}{vt2}
\fmf{dashes}{l2,vb1}
\fmf{dashes}{l6,vt1}
\fmf{dbl_plain}{r2,vb2}
\fmf{dbl_plain}{r6,vt2}
\fmfset{arrow_len}{2.2mm}
\fmf{plain,rubout=3}{xt2,yb1}
\fmf{plain}{yb2,xt1}
\fmf{phantom_arrow,rubout=2}{xt2,c}
\fmf{phantom_arrow,rubout=2}{c,yb1}
\fmf{phantom_arrow}{xt1,c}
\fmf{phantom_arrow}{c,yb2}
\fmf{fermion}{yt2,yt1}
\fmf{fermion}{xb1,xb2}
\end{fmfgraph*}}\hspace{2mm}
\\[12mm]\hspace{2mm}
\parbox{23mm}{
\begin{fmfgraph*}(23,13)
\fmfstraight
\fmfleftn{l}{7}\fmfrightn{r}{7}\fmfbottomn{b}{5}\fmftopn{t}{5}
\fmf{phantom}{l2,vb1,mb1,vb2,r2}
\fmf{phantom}{l6,vt1,mt1,vt2,r6}
\fmf{phantom}{l2,xb1,mb1,xb2,r2}
\fmf{phantom}{l6,xt1,mt1,xt2,r6}
\fmf{phantom}{l2,yb1,mb1,yb2,r2}
\fmf{phantom}{l6,yt1,mt1,yt2,r6}
\fmf{phantom}{t3,c,b3}
\fmffreeze
\fmfshift{3 right}{xb1}
\fmfshift{3 down}{xb1}
\fmfshift{3 left}{xt2}
\fmfshift{3 down}{xt2}
\fmfshift{3 right}{xt1}
\fmfshift{3 down}{xt1}
\fmfshift{3 left}{xb2}
\fmfshift{3 down}{xb2}
\fmfshift{3 right}{yb1}
\fmfshift{3 up}{yb1}
\fmfshift{3 left}{yt2}
\fmfshift{3 up}{yt2}
\fmfshift{3 right}{yt1}
\fmfshift{3 up}{yt1}
\fmfshift{3 left}{yb2}
\fmfshift{3 up}{yb2}
\fmfv{d.sh=circle,d.f=empty,d.si=8,background=(1,,.51,,.5)}{vb2}
\fmfv{d.sh=circle,d.f=empty,d.si=8,background=(.6235,,.7412,,1)}{vb1}
\fmfv{d.sh=circle,d.f=empty,d.si=8,background=(1,,.51,,.5)}{vt2}
\fmfv{d.sh=circle,d.f=empty,d.si=8,background=(.6235,,.7412,,1)}{vt1}
\fmf{dashes}{r2,vb2}
\fmf{dashes}{r6,vt2}
\fmf{dbl_plain}{l2,vb1}
\fmf{dbl_plain}{l6,vt1}
\fmfset{arrow_len}{2.2mm}
\fmf{plain,rubout=3}{yb1,xt2}
\fmf{plain}{yb2,xt1}
\fmf{phantom_arrow,rubout=2}{yb1,c}
\fmf{phantom_arrow,rubout=2}{c,xt2}
\fmf{phantom_arrow}{yb2,c}
\fmf{phantom_arrow}{c,xt1}
\fmf{fermion}{yt2,yt1}
\fmf{fermion}{xb1,xb2}
\end{fmfgraph*}}
&  \hspace{10mm} \text{\large$ 0 $}  
\end{pmatrix}
\end{align*}
\end{fmffile}   
\vspace{-3mm}

\caption{\fign{Feyn0}  Feynman diagrams making up the elements of the coupled channel $MM-D\bar D$ kernel matrix $V^{(0)}$ of \eq{V0mat}:  $V^{(0)}_{11} = 2\bGamma_1\PP_{12}\Gamma_1$,  $V^{(0)}_{12} = -2 \bGamma_1\Gamma_3$, and $V^{(0)}_{21} = -2 \bGamma_3\Gamma_1$. Solid lines with leftward (rightward) arrows represent quarks (antiquarks), dashed lines represent mesons, and double-lines represent diquarks and antidiquarks.}
\end{center}
\end{figure}

\begin{figure}[h]
\begin{center}
\begin{fmffile}{V1mat}
\begin{align*}
V^{(1)}  \hspace{1mm}&= \hspace{2mm}
\begin{pmatrix} \hspace{2mm}
\parbox{25mm}{
\begin{fmfgraph*}(25,13)
\fmfstraight
\fmfleftn{l}{7}\fmfrightn{r}{7}\fmfbottomn{b}{7}\fmftopn{t}{7}
\fmf{phantom}{l2,vb1,mb1,vb2,r2}
\fmf{phantom}{l6,vt1,mt1,vt2,r6}
\fmf{phantom}{l2,xb1,mb1,xb2,r2}
\fmf{phantom}{l6,xt1,mt1,xt2,r6}
\fmf{phantom}{l2,yb1,mb1,yb2,r2}
\fmf{phantom}{l6,yt1,mt1,yt2,r6}
\fmf{phantom}{l4,c,r4}
\fmf{phantom}{l4,clb,r4}
\fmf{phantom}{l4,clt,r4}
\fmf{phantom}{l4,crb,r4}
\fmf{phantom}{l4,crt,r4}
\fmffreeze
\fmfshift{3 right}{xb1}
\fmfshift{3 down}{xb1}
\fmfshift{3 left}{xt2}
\fmfshift{3 down}{xt2}
\fmfshift{3 right}{xt1}
\fmfshift{3 down}{xt1}
\fmfshift{3 left}{xb2}
\fmfshift{3 down}{xb2}
\fmfshift{3 right}{yb1}
\fmfshift{3 up}{yb1}
\fmfshift{3 left}{yt2}
\fmfshift{3 up}{yt2}
\fmfshift{3 right}{yt1}
\fmfshift{3 up}{yt1}
\fmfshift{3 left}{yb2}
\fmfshift{3 up}{yb2}
\fmfshift{3 left}{clb}
\fmfshift{3 down}{clb}
\fmfshift{3 left}{clt}
\fmfshift{3 up}{clt}
\fmfshift{3 right}{crb}
\fmfshift{3 down}{crb}
\fmfshift{3 right}{crt}
\fmfshift{3 up}{crt}
\fmfv{d.sh=circle,d.f=empty,d.si=8,background=(1,,.51,,.5)}{vb1}
\fmfv{d.sh=circle,d.f=empty,d.si=8,background=(1,,.51,,.5)}{vb2}
\fmfv{d.sh=circle,d.f=empty,d.si=8,background=(1,,.51,,.5)}{vt1}
\fmfv{d.sh=circle,d.f=empty,d.si=8,background=(1,,.51,,.5)}{vt2}
\fmfv{d.sh=circle,d.f=empty,d.si=8,background=(1,,.51,,.5)}{c}
\fmf{dashes}{l2,vb1}
\fmf{dashes}{l6,vt1}
\fmf{dashes}{r2,vb2}
\fmf{dashes}{r6,vt2}
\fmfset{arrow_len}{2.2mm}
\fmfi{fermion}{ vloc(__crt).. vloc(__xt2)}
\fmfi{fermion}{vloc(__xt1) ..vloc(__clt)}
\fmfi{fermion}{vloc(__yb2) ..vloc(__crb)}
\fmfi{fermion}{vloc(__clb) ..vloc(__yb1)}
\fmfi{fermion}{vloc(__yt2) .. vloc(__yt1)}
\fmfi{fermion}{vloc(__xb1) .. vloc(__xb2)}
\end{fmfgraph*}}
\hspace{0mm} + \hspace{0mm}
\parbox{25mm}{
\begin{fmfgraph*}(25,13)
\fmfstraight
\fmfleftn{l}{7}\fmfrightn{r}{7}\fmfbottomn{b}{7}\fmftopn{t}{7}
\fmf{phantom}{l2,vb1,mb1,vb2,r2}
\fmf{phantom}{l6,vt1,mt1,vt2,r6}
\fmf{phantom}{l2,xb1,mb1,xb2,r2}
\fmf{phantom}{l6,xt1,mt1,xt2,r6}
\fmf{phantom}{l2,yb1,mb1,yb2,r2}
\fmf{phantom}{l6,yt1,mt1,yt2,r6}
\fmf{phantom}{l4,c,r4}
\fmf{phantom}{l4,clb,r4}
\fmf{phantom}{l4,clt,r4}
\fmf{phantom}{l4,crb,r4}
\fmf{phantom}{l4,crt,r4}
\fmffreeze
\fmfshift{3 right}{xb1}
\fmfshift{3 down}{xb1}
\fmfshift{3 left}{xt2}
\fmfshift{3 down}{xt2}
\fmfshift{3 right}{xt1}
\fmfshift{3 down}{xt1}
\fmfshift{3 left}{xb2}
\fmfshift{3 down}{xb2}
\fmfshift{3 right}{yb1}
\fmfshift{3 up}{yb1}
\fmfshift{3 left}{yt2}
\fmfshift{3 up}{yt2}
\fmfshift{3 right}{yt1}
\fmfshift{3 up}{yt1}
\fmfshift{3 left}{yb2}
\fmfshift{3 up}{yb2}
\fmfshift{3 left}{clb}
\fmfshift{3 down}{clb}
\fmfshift{3 left}{clt}
\fmfshift{3 up}{clt}
\fmfshift{3 right}{crb}
\fmfshift{3 down}{crb}
\fmfshift{3 right}{crt}
\fmfshift{3 up}{crt}
\fmfv{d.sh=circle,d.f=empty,d.si=8,background=(1,,.51,,.5)}{vb1}
\fmfv{d.sh=circle,d.f=empty,d.si=8,background=(1,,.51,,.5)}{vb2}
\fmfv{d.sh=circle,d.f=empty,d.si=8,background=(1,,.51,,.5)}{vt1}
\fmfv{d.sh=circle,d.f=empty,d.si=8,background=(1,,.51,,.5)}{vt2}
\fmfv{d.sh=circle,d.f=empty,d.si=8,background=(.6235,,.7412,,1)}{c}
\fmf{dashes}{l2,vb1}
\fmf{dashes}{l6,vt1}
\fmf{dashes}{r2,vb2}
\fmf{dashes}{r6,vt2}
\fmfset{arrow_len}{2.2mm}
\fmfi{fermion}{vloc(__crt) .. vloc(__xt2)}
\fmfi{fermion}{vloc(__xt1) ..vloc(__clt)}
\fmfi{fermion}{vloc(__crb) ..vloc(__yb2)}
\fmfi{fermion}{vloc(__yb1) ..vloc(__clb)}
\fmfi{fermion}{vloc(__yt2) .. vloc(__yt1)}
\fmfi{fermion}{vloc(__xb2) .. vloc(__xb1)}
\end{fmfgraph*}} \hspace{10mm} &
\parbox{25mm}{
\begin{fmfgraph*}(25,13)
\fmfstraight
\fmfleftn{l}{7}\fmfrightn{r}{7}\fmfbottomn{b}{7}\fmftopn{t}{7}
\fmf{phantom}{l2,vb1,mb1,vb2,r2}
\fmf{phantom}{l6,vt1,mt1,vt2,r6}
\fmf{phantom}{l2,xb1,mb1,xb2,r2}
\fmf{phantom}{l6,xt1,mt1,xt2,r6}
\fmf{phantom}{l2,yb1,mb1,yb2,r2}
\fmf{phantom}{l6,yt1,mt1,yt2,r6}
\fmf{phantom}{l4,c,r4}
\fmf{phantom}{l4,clb,r4}
\fmf{phantom}{l4,clt,r4}
\fmf{phantom}{l4,crb,r4}
\fmf{phantom}{l4,crt,r4}
\fmffreeze
\fmfshift{3 right}{xb1}
\fmfshift{3 down}{xb1}
\fmfshift{3 left}{xt2}
\fmfshift{3 down}{xt2}
\fmfshift{3 right}{xt1}
\fmfshift{3 down}{xt1}
\fmfshift{3 left}{xb2}
\fmfshift{3 down}{xb2}
\fmfshift{3 right}{yb1}
\fmfshift{3 up}{yb1}
\fmfshift{3 left}{yt2}
\fmfshift{3 up}{yt2}
\fmfshift{3 right}{yt1}
\fmfshift{3 up}{yt1}
\fmfshift{3 left}{yb2}
\fmfshift{3 up}{yb2}
\fmfshift{3 left}{clb}
\fmfshift{3 down}{clb}
\fmfshift{3 left}{clt}
\fmfshift{3 up}{clt}
\fmfshift{3 right}{crb}
\fmfshift{3 down}{crb}
\fmfshift{3 right}{crt}
\fmfshift{3 up}{crt}
\fmfv{d.sh=circle,d.f=empty,d.si=8,background=(1,,.51,,.5)}{vb1}
\fmfv{d.sh=circle,d.f=empty,d.si=8,background=(.6235,,.7412,,1)}{vb2}
\fmfv{d.sh=circle,d.f=empty,d.si=8,background=(1,,.51,,.5)}{vt1}
\fmfv{d.sh=circle,d.f=empty,d.si=8,background=(.6235,,.7412,,1)}{vt2}
\fmfv{d.sh=circle,d.f=empty,d.si=8,background=(1,,.51,,.5)}{c}
\fmf{dashes}{l2,vb1}
\fmf{dashes}{l6,vt1}
\fmf{dbl_plain}{r2,vb2}
\fmf{dbl_plain}{r6,vt2}
\fmfset{arrow_len}{2.2mm}
\fmfi{fermion}{ vloc(__xt2).. vloc(__crt)}
\fmfi{fermion}{vloc(__xt1) ..vloc(__clt)}
\fmfi{fermion}{vloc(__crb).. vloc(__yb2)}
\fmfi{fermion}{vloc(__clb).. vloc(__yb1)}
\fmfi{fermion}{vloc(__yt2) .. vloc(__yt1)}
\fmfi{fermion}{vloc(__xb1) .. vloc(__xb2)}
\end{fmfgraph*}}\hspace{2mm}
\\[12mm]
\hspace{-8mm}
\parbox{25mm}{
\begin{fmfgraph*}(25,13)
\fmfstraight
\fmfleftn{l}{7}\fmfrightn{r}{7}\fmfbottomn{b}{7}\fmftopn{t}{7}
\fmf{phantom}{l2,vb1,mb1,vb2,r2}
\fmf{phantom}{l6,vt1,mt1,vt2,r6}
\fmf{phantom}{l2,xb1,mb1,xb2,r2}
\fmf{phantom}{l6,xt1,mt1,xt2,r6}
\fmf{phantom}{l2,yb1,mb1,yb2,r2}
\fmf{phantom}{l6,yt1,mt1,yt2,r6}
\fmf{phantom}{l4,c,r4}
\fmf{phantom}{l4,clb,r4}
\fmf{phantom}{l4,clt,r4}
\fmf{phantom}{l4,crb,r4}
\fmf{phantom}{l4,crt,r4}
\fmffreeze
\fmfshift{3 right}{xb1}
\fmfshift{3 down}{xb1}
\fmfshift{3 left}{xt2}
\fmfshift{3 down}{xt2}
\fmfshift{3 right}{xt1}
\fmfshift{3 down}{xt1}
\fmfshift{3 left}{xb2}
\fmfshift{3 down}{xb2}
\fmfshift{3 right}{yb1}
\fmfshift{3 up}{yb1}
\fmfshift{3 left}{yt2}
\fmfshift{3 up}{yt2}
\fmfshift{3 right}{yt1}
\fmfshift{3 up}{yt1}
\fmfshift{3 left}{yb2}
\fmfshift{3 up}{yb2}
\fmfshift{3 left}{clb}
\fmfshift{3 down}{clb}
\fmfshift{3 left}{clt}
\fmfshift{3 up}{clt}
\fmfshift{3 right}{crb}
\fmfshift{3 down}{crb}
\fmfshift{3 right}{crt}
\fmfshift{3 up}{crt}
\fmfv{d.sh=circle,d.f=empty,d.si=8,background=(.6235,,.7412,,1)}{vb1}
\fmfv{d.sh=circle,d.f=empty,d.si=8,background=(1,,.51,,.5)}{vb2}
\fmfv{d.sh=circle,d.f=empty,d.si=8,background=(.6235,,.7412,,1)}{vt1}
\fmfv{d.sh=circle,d.f=empty,d.si=8,background=(1,,.51,,.5)}{vt2}
\fmfv{d.sh=circle,d.f=empty,d.si=8,background=(1,,.51,,.5)}{c}
\fmf{dbl_plain}{l2,vb1}
\fmf{dbl_plain}{l6,vt1}
\fmf{dashes}{r2,vb2}
\fmf{dashes}{r6,vt2}
\fmfset{arrow_len}{2.2mm}
\fmfi{fermion}{ vloc(__crt).. vloc(__xt2)}
\fmfi{fermion}{vloc(__clt) ..vloc(__xt1)}
\fmfi{fermion}{vloc(__yb2)..vloc(__crb) }
\fmfi{fermion}{ vloc(__yb1)..vloc(__clb)}
\fmfi{fermion}{vloc(__yt2) .. vloc(__yt1)}
\fmfi{fermion}{vloc(__xb1) .. vloc(__xb2)}
\end{fmfgraph*}}
&
\parbox{23mm}{
\begin{fmfgraph*}(23,13)
\fmfstraight
\fmfleftn{l}{7}\fmfrightn{r}{7}\fmfbottomn{b}{7}\fmftopn{t}{7}
\fmf{phantom}{l2,vb1,mb1,vb2,r2}
\fmf{phantom}{l6,vt1,mt1,vt2,r6}
\fmf{phantom}{l2,xb1,mb1,xb2,r2}
\fmf{phantom}{l6,xt1,mt1,xt2,r6}
\fmf{phantom}{l2,yb1,mb1,yb2,r2}
\fmf{phantom}{l6,yt1,mt1,yt2,r6}
\fmf{phantom}{l4,c,r4}
\fmf{phantom}{l4,clb,r4}
\fmf{phantom}{l4,clt,r4}
\fmf{phantom}{l4,crb,r4}
\fmf{phantom}{l4,crt,r4}
\fmffreeze
\fmfshift{3 right}{xb1}
\fmfshift{3 down}{xb1}
\fmfshift{3 left}{xt2}
\fmfshift{3 down}{xt2}
\fmfshift{3 right}{xt1}
\fmfshift{3 down}{xt1}
\fmfshift{3 left}{xb2}
\fmfshift{3 down}{xb2}
\fmfshift{3 right}{yb1}
\fmfshift{3 up}{yb1}
\fmfshift{3 left}{yt2}
\fmfshift{3 up}{yt2}
\fmfshift{3 right}{yt1}
\fmfshift{3 up}{yt1}
\fmfshift{3 left}{yb2}
\fmfshift{3 up}{yb2}
\fmfshift{3 left}{clb}
\fmfshift{3 down}{clb}
\fmfshift{3 left}{clt}
\fmfshift{3 up}{clt}
\fmfshift{3 right}{crb}
\fmfshift{3 down}{crb}
\fmfshift{3 right}{crt}
\fmfshift{3 up}{crt}
\fmfv{d.sh=circle,d.f=empty,d.si=8,background=(.6235,,.7412,,1)}{vb1}
\fmfv{d.sh=circle,d.f=empty,d.si=8,background=(.6235,,.7412,,1)}{vb2}
\fmfv{d.sh=circle,d.f=empty,d.si=8,background=(.6235,,.7412,,1)}{vt1}
\fmfv{d.sh=circle,d.f=empty,d.si=8,background=(.6235,,.7412,,1)}{vt2}
\fmfv{d.sh=circle,d.f=empty,d.si=8,background=(1,,.51,,.5)}{c}
\fmf{dbl_plain}{l2,vb1}
\fmf{dbl_plain}{l6,vt1}
\fmf{dbl_plain}{r2,vb2}
\fmf{dbl_plain}{r6,vt2}
\fmfset{arrow_len}{2.2mm}
\fmfi{fermion}{vloc(__xt2) .. vloc(__crt)}
\fmfi{fermion}{ vloc(__clt)..vloc(__xt1)}
\fmfi{fermion}{vloc(__crb) ..vloc(__yb2)}
\fmfi{fermion}{vloc(__yb1) ..vloc(__clb)}
\fmfi{fermion}{vloc(__yt2) .. vloc(__yt1)}
\fmfi{fermion}{vloc(__xb1) .. vloc(__xb2)}
\end{fmfgraph*}} \hspace{2mm}
\end{pmatrix}
\end{align*}
\end{fmffile}   
\vspace{-3mm}

\caption{\fign{Feyn1}  Feynman diagrams making up the elements of the coupled channel $MM-D\bar D$ kernel matrix $V^{(1)}$ of \eq{V1mat}: $V^{(1)}_{11} = -2\left( \bGamma_1\PP_{12}T^+_1\PP_{12}\Gamma_1 +  2\bGamma_1 T^+_3\Gamma_1\right)$,  $V^{(1)}_{12} =  2\bGamma_1\PP_{12}T^+_1 \Gamma_3$, $V^{(1)}_{21} = 2 \bGamma_3 T^+_1\PP_{12}\Gamma_1$, and $V^{(1)}_{22} = -4\bGamma_3T^+_1\Gamma_3$.  Solid lines with leftward (rightward) arrows represent quarks (antiquarks), dashed lines represent mesons, and double-lines represent diquarks and antidiquarks.}
\end{center}
\end{figure}

In the present work, we extend the unified tetraquark equations of Ref.\ \cite{Kvinikhidze:2023djp}, namely \eq{eq}, to include all contributions that have been neglected in the derivation of these equations. In particular, we explicitly include $q\q$ annihilation within the same formalism as used in \eq{eq}, and at the same time, take into account all possible other contributions (e.g. three-body forces,  non-pole contributions to the t matrices within the products $T_a T_{a'}$, correction terms to overcounted or undercounted contributions, etc.) through the introduction of a  $q\q$  amplitude $\Delta$ that is defined to consist of all corrections needed to restore the exactness of the model. That this is possible has been demonstrated by us in Ref.  \cite{Kvinikhidze:2021kzu} on the example where the first two terms of \eq{Taa} have been neglected. By following the same approach as \cite{Kvinikhidze:2021kzu}, but retaining the first two terms of \eq{Taa},
we obtain equations of the same form as the unified tetraquark equations of \eq{eq}, but that include $q\q$ annihilation in the way prescribed by quantum field theory, and that are otherwise exact due to the inclusion of the $\Delta$ amplitude. Similarly to \eq{eq}, these equation can be expressed in matrix form as
\be
\varphi = \left[ \left( {\cal V}^{(0)} + {\cal V}^{(1)}+\dots\right) + {\cal W} \right] {\cal D} \varphi    \eqn{eq3}
\ee
where $\varphi$ is a rank-3 column of tetraquark form factors, ${\cal D}$ is a diagonal matrix consisting of the two-meson propagator $MM$, the diquark-antidiquark propagator $D\D$, and the $q\q$ propagator $Q\bar Q$, namely,
\be
\varphi = \begin{pmatrix} \phi_M \\  \phi_D \\[1mm] \Gamma^* \end{pmatrix}
=
\begin{fmffile}{Phii}
\begin{pmatrix} 
\parbox{10mm}{
\begin{fmfgraph*}(10,13)
\fmfset{dash_len}{2mm}
\fmfstraight
\fmfleftn{l}{7}\fmfrightn{r}{7}\fmfbottomn{b}{6}\fmftopn{t}{6}
\fmf{phantom}{l2,mb2,r2}
\fmf{phantom}{l6,mt2,r6}
\fmffreeze
\fmf{phantom}{r2,phi,r6}
\fmffreeze
\fmf{dashes}{phi,l2}
\fmf{dashes}{phi,l6}
\fmfv{d.sh=circle,d.f=empty,d.si=22,label=$\hspace{-4.mm}\phi_{\!M}$,background=(1,,.51,,.5)}{phi}
\end{fmfgraph*}} \hspace{4mm} \\
\parbox{10mm}{
\begin{fmfgraph*}(10,13)
\fmfstraight
\fmfleftn{l}{7}\fmfrightn{r}{7}\fmfbottomn{b}{6}\fmftopn{t}{6}
\fmf{phantom}{l2,mb2,r2}
\fmf{phantom}{l6,mt2,r6}
\fmffreeze
\fmf{phantom}{r2,phi,r6}
\fmffreeze
\fmf{dbl_plain}{phi,l2}
\fmf{dbl_plain}{phi,l6}
\fmfv{d.sh=circle,d.f=empty,d.si=22,label=$\hspace{-4.mm}\phi_{\!D}$,background=(.6235,,.7412,,1)}{phi}
\end{fmfgraph*}}\hspace{3mm} \\
\parbox{10mm}{
\begin{fmfgraph*}(10,13)
\fmfstraight
\fmfleftn{l}{7}\fmfrightn{r}{7}\fmfbottomn{b}{3}\fmftopn{t}{3}
\fmf{phantom}{l2,mb2,r2}
\fmf{phantom}{l6,mt2,r6}
\fmffreeze
\fmf{phantom}{r2,phi,r6}
\fmffreeze
\fmfset{arrow_len}{2.2mm}
\fmf{fermion}{l2,phi}
\fmf{fermion}{phi,l6}
\fmfv{d.sh=circle,d.f=empty,d.si=22,label=$\hspace{-4.mm}\Gamma^*$,background=(1,,.9333,,.333)}{phi}
\end{fmfgraph*}}\hspace{3mm}
\end{pmatrix}
\end{fmffile},
\hspace{1cm} {\cal D}= \begin{pmatrix} \frac{1}{2} M M  & 0 & 0\\ 0 & D\D & 0\\
0 & 0 & Q\bar Q \end{pmatrix},
\ee
where $\Phi_M$, $\Phi_D$, and $\Gamma^*$  are form factors (together with their graphical representations) that couple the tetraquark to $MM$, $D\D$ and $q\q$ states, respectively. The kernel of \eq{eq3}  (term in square brackets) consists of two parts: (i) the kernel ${\cal V}$, defined formally as the sum of the infinite series
\be
{\cal V} \equiv  {\cal V}^{(0)} + {\cal V}^{(1)} + \dots    \eqn{calV}
\ee
which is the kernel of the corresponding equation without $q\q$ annihilation, i.e.\ \eq{eq}, but expressed in rank-3 form as illustrated for the first two terms ${\cal V}^{(0)}$ and ${\cal V}^{(1)}$  in \fig{Feyn30} and \fig{Feyn31}, respectively, and (ii) the kernel ${\cal W}$, illustrated in \fig{FeynW}, which consists of $q\q$-irreducible amplitudes $\bar N_M$, $\bar N_D$, $N_M$, and $N_D$ connecting $MM$ and $D\D$ states to $q\q$ states [as described in the figure caption of \fig{FeynW}], and the aforementioned $q\q$ elastic scattering amplitude $\Delta$.

A feature of the tetraquark description as modelled by \eq{eq3}, is that it is formally exact no matter what model or approximations are used in calculating the kernel ${\cal V}$, or what model is used for the ``$N$ amplitudes" $\bar N_M$, $\bar N_D$, $N_M$, and $N_D$. This is due to the fact that the $\Delta$ appearing in ${\cal W}$, by definition, consists of compensating terms that keep the tetraquark description exact. 
To illustrate this point, let us consider the model where ${\cal V} = {\cal V}^{(0)}$, corresponding to the Giessen model, but where $q\q$ annihilation is taken into account through the inclusion of the kernel ${\cal W}$ where the $N$ amplitudes are modelled as in \fig{FeynW}. We can then formally adjust the $\Delta$ amplitude to the amplitude $\Delta_0$, with the resulting kernel ${\cal W}$ being denoted by ${\cal W}_0$, such that the tetraquark equation
\be
\varphi = \left[ {\cal V}^{(0)}  + {\cal W}_0 \right] {\cal D} \varphi    \eqn{eq30}
\ee
is exact; that is, having the same solution $\varphi$ as \eq{eq3}. {That this is the case can be seen from the discussion in Sec.\ 2.1 below, where we show that the three equations represented by \eq{eq3} are equivalent to the single equation for the $\text{tetraquark}\rightarrow q\q$ form factor $\Gamma^*$,  \eq{2b-tetr}, whose $q\q$ kernel $K^{(2)}$, given by \eq{K2},  can be made exact  by adjusting  $\Delta$, no matter what model is chosen for the $N$ amplitudes or potential $V$.} This means that $\Delta_0$ contains within it information about ${\cal V}^{(1)}$ and all the rest of the contributions not explicitly appearing in \eq{eq30}. Similarly, we can consider the model where ${\cal V} = {\cal V}^{(0)} +  {\cal V}^{(1)}$,  and then adjust the $\Delta$ amplitude to the amplitude $\Delta_1$, with the resulting kernel ${\cal W}$ being denoted by ${\cal W}_1$, such that the tetraquark equation
\be
\varphi = \left[ {\cal V}^{(0)}  +{\cal V}^{(1)}  + {\cal W}_1 \right] {\cal D} \varphi    \eqn{eq31}
\ee
is exact. The amplitude $\Delta_1$ then contains within it information about ${\cal V}^{(2)}$ and all the rest of the contributions not explicitly appearing in \eq{eq31}. Equation (\ref{eq31}) is then the exact tetraquark equation that includes $q\q$ annihilation into the unified description of the Giessen and Moscow models of the tetraquark as described by Eq.\ (61) of \cite{Kvinikhidze:2023djp}.  It is clear from these examples that the tetraquark equations, \eq{eq3}, have the generality to unify a large variety of other possible tetraquark models. {Moreover, the same equations can be applied generally to four-body systems that couple to 2-body channels, as for example the two-electron plus two-positron system, the two-nucleon plus two-antinucleon system, etc.}

\begin{figure}[t]
\begin{center}
\begin{fmffile}{V30mat}
\begin{align*}
{\cal V}^{(0)} \hspace{1mm} & = \hspace{2mm}
\begin{pmatrix}  \hspace{2mm}
\parbox{23mm}{
\begin{fmfgraph*}(23,13)
\fmfstraight
\fmfleftn{l}{7}\fmfrightn{r}{7}\fmfbottomn{b}{5}\fmftopn{t}{5}
\fmf{phantom}{l2,vb1,mb1,vb2,r2}
\fmf{phantom}{l6,vt1,mt1,vt2,r6}
\fmf{phantom}{l2,xb1,mb1,xb2,r2}
\fmf{phantom}{l6,xt1,mt1,xt2,r6}
\fmf{phantom}{l2,yb1,mb1,yb2,r2}
\fmf{phantom}{l6,yt1,mt1,yt2,r6}
\fmf{phantom}{t3,c,b3}
\fmffreeze
\fmfshift{3 right}{xb1}
\fmfshift{3 down}{xb1}
\fmfshift{3 left}{xt2}
\fmfshift{3 down}{xt2}
\fmfshift{3 right}{xt1}
\fmfshift{3 down}{xt1}
\fmfshift{3 left}{xb2}
\fmfshift{3 down}{xb2}
\fmfshift{3 right}{yb1}
\fmfshift{3 up}{yb1}
\fmfshift{3 left}{yt2}
\fmfshift{3 up}{yt2}
\fmfshift{3 right}{yt1}
\fmfshift{3 up}{yt1}
\fmfshift{3 left}{yb2}
\fmfshift{3 up}{yb2}
\fmfv{d.sh=circle,d.f=empty,d.si=8,background=(1,,.51,,.5)}{vb1}
\fmfv{d.sh=circle,d.f=empty,d.si=8,background=(1,,.51,,.5)}{vb2}
\fmfv{d.sh=circle,d.f=empty,d.si=8,background=(1,,.51,,.5)}{vt1}
\fmfv{d.sh=circle,d.f=empty,d.si=8,background=(1,,.51,,.5)}{vt2}
\fmf{dashes}{l2,vb1}
\fmf{dashes}{l6,vt1}
\fmf{dashes}{r2,vb2}
\fmf{dashes}{r6,vt2}
\fmfset{arrow_len}{2.2mm}
\fmf{plain,rubout=3}{yb1,xt2}
\fmf{plain}{xt1,yb2}
\fmf{phantom_arrow}{xt1,c}
\fmf{phantom_arrow}{c,yb2}
\fmf{phantom_arrow,rubout=2}{yb1,c}
\fmf{phantom_arrow,rubout=2}{c,xt2}
\fmf{fermion}{yt2,yt1}
\fmf{fermion}{xb2,xb1}
\end{fmfgraph*}}
&  \hspace{8mm}
\parbox{23mm}{
\begin{fmfgraph*}(23,13)
\fmfstraight
\fmfleftn{l}{7}\fmfrightn{r}{7}\fmfbottomn{b}{5}\fmftopn{t}{5}
\fmf{phantom}{l2,vb1,mb1,vb2,r2}
\fmf{phantom}{l6,vt1,mt1,vt2,r6}
\fmf{phantom}{l2,xb1,mb1,xb2,r2}
\fmf{phantom}{l6,xt1,mt1,xt2,r6}
\fmf{phantom}{l2,yb1,mb1,yb2,r2}
\fmf{phantom}{l6,yt1,mt1,yt2,r6}
\fmf{phantom}{t3,c,b3}
\fmffreeze
\fmfshift{3 right}{xb1}
\fmfshift{3 down}{xb1}
\fmfshift{3 left}{xt2}
\fmfshift{3 down}{xt2}
\fmfshift{3 right}{xt1}
\fmfshift{3 down}{xt1}
\fmfshift{3 left}{xb2}
\fmfshift{3 down}{xb2}
\fmfshift{3 right}{yb1}
\fmfshift{3 up}{yb1}
\fmfshift{3 left}{yt2}
\fmfshift{3 up}{yt2}
\fmfshift{3 right}{yt1}
\fmfshift{3 up}{yt1}
\fmfshift{3 left}{yb2}
\fmfshift{3 up}{yb2}
\fmfv{d.sh=circle,d.f=empty,d.si=8,background=(1,,.51,,.5)}{vb1}
\fmfv{d.sh=circle,d.f=empty,d.si=8,background=(.6235,,.7412,,1)}{vb2}
\fmfv{d.sh=circle,d.f=empty,d.si=8,background=(1,,.51,,.5)}{vt1}
\fmfv{d.sh=circle,d.f=empty,d.si=8,background=(.6235,,.7412,,1)}{vt2}
\fmf{dashes}{l2,vb1}
\fmf{dashes}{l6,vt1}
\fmf{dbl_plain}{r2,vb2}
\fmf{dbl_plain}{r6,vt2}
\fmfset{arrow_len}{2.2mm}
\fmf{plain,rubout=3}{xt2,yb1}
\fmf{plain}{yb2,xt1}
\fmf{phantom_arrow,rubout=2}{xt2,c}
\fmf{phantom_arrow,rubout=2}{c,yb1}
\fmf{phantom_arrow}{xt1,c}
\fmf{phantom_arrow}{c,yb2}
\fmf{fermion}{yt2,yt1}
\fmf{fermion}{xb1,xb2}
\end{fmfgraph*}}
&  \hspace{6mm}
 \text{\large$ 0 $}
\hspace{2mm}
\\[8mm]\hspace{2mm}
\parbox{23mm}{
\begin{fmfgraph*}(23,13)
\fmfstraight
\fmfleftn{l}{7}\fmfrightn{r}{7}\fmfbottomn{b}{5}\fmftopn{t}{5}
\fmf{phantom}{l2,vb1,mb1,vb2,r2}
\fmf{phantom}{l6,vt1,mt1,vt2,r6}
\fmf{phantom}{l2,xb1,mb1,xb2,r2}
\fmf{phantom}{l6,xt1,mt1,xt2,r6}
\fmf{phantom}{l2,yb1,mb1,yb2,r2}
\fmf{phantom}{l6,yt1,mt1,yt2,r6}
\fmf{phantom}{t3,c,b3}
\fmffreeze
\fmfshift{3 right}{xb1}
\fmfshift{3 down}{xb1}
\fmfshift{3 left}{xt2}
\fmfshift{3 down}{xt2}
\fmfshift{3 right}{xt1}
\fmfshift{3 down}{xt1}
\fmfshift{3 left}{xb2}
\fmfshift{3 down}{xb2}
\fmfshift{3 right}{yb1}
\fmfshift{3 up}{yb1}
\fmfshift{3 left}{yt2}
\fmfshift{3 up}{yt2}
\fmfshift{3 right}{yt1}
\fmfshift{3 up}{yt1}
\fmfshift{3 left}{yb2}
\fmfshift{3 up}{yb2}
\fmfv{d.sh=circle,d.f=empty,d.si=8,background=(1,,.51,,.5)}{vb2}
\fmfv{d.sh=circle,d.f=empty,d.si=8,background=(.6235,,.7412,,1)}{vb1}
\fmfv{d.sh=circle,d.f=empty,d.si=8,background=(1,,.51,,.5)}{vt2}
\fmfv{d.sh=circle,d.f=empty,d.si=8,background=(.6235,,.7412,,1)}{vt1}
\fmf{dashes}{r2,vb2}
\fmf{dashes}{r6,vt2}
\fmf{dbl_plain}{l2,vb1}
\fmf{dbl_plain}{l6,vt1}
\fmfset{arrow_len}{2.2mm}
\fmf{plain,rubout=3}{yb1,xt2}
\fmf{plain}{yb2,xt1}
\fmf{phantom_arrow,rubout=2}{yb1,c}
\fmf{phantom_arrow,rubout=2}{c,xt2}
\fmf{phantom_arrow}{yb2,c}
\fmf{phantom_arrow}{c,xt1}
\fmf{fermion}{yt2,yt1}
\fmf{fermion}{xb1,xb2}
\end{fmfgraph*}}
&  \hspace{8mm} \text{\large$ 0 $}
& \hspace{6mm}
 \text{\large$ 0 $}
 \hspace{2mm}
\\[9mm]\hspace{2mm}
 \text{\large$ 0 $} &
\hspace{8mm}
 \text{\large$ 0 $}&\hspace{4mm}
 \text{\large$ 0 $}\end{pmatrix}
\end{align*}
\end{fmffile}   
\vspace{-3mm}

\caption{\fign{Feyn30}  The kernel matrix ${\cal V}^{(0)}$ defined as the rank-3 version of $V^{(0)}$, illustrated in \fig{Feyn0}.}
\end{center}
\end{figure}

\begin{figure}[t]
\begin{center}
\begin{fmffile}{V31mat}
\begin{align*}
{\cal V}^{(1)}  \hspace{1mm}&= \hspace{2mm}
\begin{pmatrix} \hspace{2mm}
\parbox{23mm}{
\begin{fmfgraph*}(23,13.)
\fmfstraight
\fmfleftn{l}{7}\fmfrightn{r}{7}\fmfbottomn{b}{7}\fmftopn{t}{7}
\fmf{phantom}{l2,vb1,mb1,vb2,r2}
\fmf{phantom}{l6,vt1,mt1,vt2,r6}
\fmf{phantom}{l2,xb1,mb1,xb2,r2}
\fmf{phantom}{l6,xt1,mt1,xt2,r6}
\fmf{phantom}{l2,yb1,mb1,yb2,r2}
\fmf{phantom}{l6,yt1,mt1,yt2,r6}
\fmf{phantom}{l4,c,r4}
\fmf{phantom}{l4,clb,r4}
\fmf{phantom}{l4,clt,r4}
\fmf{phantom}{l4,crb,r4}
\fmf{phantom}{l4,crt,r4}
\fmffreeze
\fmfshift{3 right}{xb1}
\fmfshift{3 down}{xb1}
\fmfshift{3 left}{xt2}
\fmfshift{3 down}{xt2}
\fmfshift{3 right}{xt1}
\fmfshift{3 down}{xt1}
\fmfshift{3 left}{xb2}
\fmfshift{3 down}{xb2}
\fmfshift{3 right}{yb1}
\fmfshift{3 up}{yb1}
\fmfshift{3 left}{yt2}
\fmfshift{3 up}{yt2}
\fmfshift{3 right}{yt1}
\fmfshift{3 up}{yt1}
\fmfshift{3 left}{yb2}
\fmfshift{3 up}{yb2}
\fmfshift{3 left}{clb}
\fmfshift{3 down}{clb}
\fmfshift{3 left}{clt}
\fmfshift{3 up}{clt}
\fmfshift{3 right}{crb}
\fmfshift{3 down}{crb}
\fmfshift{3 right}{crt}
\fmfshift{3 up}{crt}
\fmfv{d.sh=circle,d.f=empty,d.si=8,background=(1,,.51,,.5)}{vb1}
\fmfv{d.sh=circle,d.f=empty,d.si=8,background=(1,,.51,,.5)}{vb2}
\fmfv{d.sh=circle,d.f=empty,d.si=8,background=(1,,.51,,.5)}{vt1}
\fmfv{d.sh=circle,d.f=empty,d.si=8,background=(1,,.51,,.5)}{vt2}
\fmfv{d.sh=circle,d.f=empty,d.si=8,background=(1,,.51,,.5)}{c}
\fmf{dashes}{l2,vb1}
\fmf{dashes}{l6,vt1}
\fmf{dashes}{r2,vb2}
\fmf{dashes}{r6,vt2}
\fmfset{arrow_len}{2.2mm}
\fmfi{fermion}{ vloc(__crt).. vloc(__xt2)}
\fmfi{fermion}{vloc(__xt1) ..vloc(__clt)}
\fmfi{fermion}{vloc(__yb2) ..vloc(__crb)}
\fmfi{fermion}{vloc(__clb) ..vloc(__yb1)}
\fmfi{fermion}{vloc(__yt2) .. vloc(__yt1)}
\fmfi{fermion}{vloc(__xb1) .. vloc(__xb2)}
\end{fmfgraph*}}
\hspace{0mm} + \hspace{0mm}
\parbox{23mm}{
\begin{fmfgraph*}(23,13.)
\fmfstraight
\fmfleftn{l}{7}\fmfrightn{r}{7}\fmfbottomn{b}{7}\fmftopn{t}{7}
\fmf{phantom}{l2,vb1,mb1,vb2,r2}
\fmf{phantom}{l6,vt1,mt1,vt2,r6}
\fmf{phantom}{l2,xb1,mb1,xb2,r2}
\fmf{phantom}{l6,xt1,mt1,xt2,r6}
\fmf{phantom}{l2,yb1,mb1,yb2,r2}
\fmf{phantom}{l6,yt1,mt1,yt2,r6}
\fmf{phantom}{l4,c,r4}
\fmf{phantom}{l4,clb,r4}
\fmf{phantom}{l4,clt,r4}
\fmf{phantom}{l4,crb,r4}
\fmf{phantom}{l4,crt,r4}
\fmffreeze
\fmfshift{3 right}{xb1}
\fmfshift{3 down}{xb1}
\fmfshift{3 left}{xt2}
\fmfshift{3 down}{xt2}
\fmfshift{3 right}{xt1}
\fmfshift{3 down}{xt1}
\fmfshift{3 left}{xb2}
\fmfshift{3 down}{xb2}
\fmfshift{3 right}{yb1}
\fmfshift{3 up}{yb1}
\fmfshift{3 left}{yt2}
\fmfshift{3 up}{yt2}
\fmfshift{3 right}{yt1}
\fmfshift{3 up}{yt1}
\fmfshift{3 left}{yb2}
\fmfshift{3 up}{yb2}
\fmfshift{3 left}{clb}
\fmfshift{3 down}{clb}
\fmfshift{3 left}{clt}
\fmfshift{3 up}{clt}
\fmfshift{3 right}{crb}
\fmfshift{3 down}{crb}
\fmfshift{3 right}{crt}
\fmfshift{3 up}{crt}
\fmfv{d.sh=circle,d.f=empty,d.si=8,background=(1,,.51,,.5)}{vb1}
\fmfv{d.sh=circle,d.f=empty,d.si=8,background=(1,,.51,,.5)}{vb2}
\fmfv{d.sh=circle,d.f=empty,d.si=8,background=(1,,.51,,.5)}{vt1}
\fmfv{d.sh=circle,d.f=empty,d.si=8,background=(1,,.51,,.5)}{vt2}
\fmfv{d.sh=circle,d.f=empty,d.si=8,background=(.6235,,.7412,,1)}{c}
\fmf{dashes}{l2,vb1}
\fmf{dashes}{l6,vt1}
\fmf{dashes}{r2,vb2}
\fmf{dashes}{r6,vt2}
\fmfset{arrow_len}{2.2mm}
\fmfi{fermion}{vloc(__crt) .. vloc(__xt2)}
\fmfi{fermion}{vloc(__xt1) ..vloc(__clt)}
\fmfi{fermion}{vloc(__crb) ..vloc(__yb2)}
\fmfi{fermion}{vloc(__yb1) ..vloc(__clb)}
\fmfi{fermion}{vloc(__yt2) .. vloc(__yt1)}
\fmfi{fermion}{vloc(__xb2) .. vloc(__xb1)}
\end{fmfgraph*}} \hspace{8mm} &
\parbox{23mm}{
\begin{fmfgraph*}(23,13.)
\fmfstraight
\fmfleftn{l}{7}\fmfrightn{r}{7}\fmfbottomn{b}{7}\fmftopn{t}{7}
\fmf{phantom}{l2,vb1,mb1,vb2,r2}
\fmf{phantom}{l6,vt1,mt1,vt2,r6}
\fmf{phantom}{l2,xb1,mb1,xb2,r2}
\fmf{phantom}{l6,xt1,mt1,xt2,r6}
\fmf{phantom}{l2,yb1,mb1,yb2,r2}
\fmf{phantom}{l6,yt1,mt1,yt2,r6}
\fmf{phantom}{l4,c,r4}
\fmf{phantom}{l4,clb,r4}
\fmf{phantom}{l4,clt,r4}
\fmf{phantom}{l4,crb,r4}
\fmf{phantom}{l4,crt,r4}
\fmffreeze
\fmfshift{3 right}{xb1}
\fmfshift{3 down}{xb1}
\fmfshift{3 left}{xt2}
\fmfshift{3 down}{xt2}
\fmfshift{3 right}{xt1}
\fmfshift{3 down}{xt1}
\fmfshift{3 left}{xb2}
\fmfshift{3 down}{xb2}
\fmfshift{3 right}{yb1}
\fmfshift{3 up}{yb1}
\fmfshift{3 left}{yt2}
\fmfshift{3 up}{yt2}
\fmfshift{3 right}{yt1}
\fmfshift{3 up}{yt1}
\fmfshift{3 left}{yb2}
\fmfshift{3 up}{yb2}
\fmfshift{3 left}{clb}
\fmfshift{3 down}{clb}
\fmfshift{3 left}{clt}
\fmfshift{3 up}{clt}
\fmfshift{3 right}{crb}
\fmfshift{3 down}{crb}
\fmfshift{3 right}{crt}
\fmfshift{3 up}{crt}
\fmfv{d.sh=circle,d.f=empty,d.si=8,background=(1,,.51,,.5)}{vb1}
\fmfv{d.sh=circle,d.f=empty,d.si=8,background=(.6235,,.7412,,1)}{vb2}
\fmfv{d.sh=circle,d.f=empty,d.si=8,background=(1,,.51,,.5)}{vt1}
\fmfv{d.sh=circle,d.f=empty,d.si=8,background=(.6235,,.7412,,1)}{vt2}
\fmfv{d.sh=circle,d.f=empty,d.si=8,background=(1,,.51,,.5)}{c}
\fmf{dashes}{l2,vb1}
\fmf{dashes}{l6,vt1}
\fmf{dbl_plain}{r2,vb2}
\fmf{dbl_plain}{r6,vt2}
\fmfset{arrow_len}{2.2mm}
\fmfi{fermion}{ vloc(__xt2).. vloc(__crt)}
\fmfi{fermion}{vloc(__xt1) ..vloc(__clt)}
\fmfi{fermion}{vloc(__crb).. vloc(__yb2)}
\fmfi{fermion}{vloc(__clb).. vloc(__yb1)}
\fmfi{fermion}{vloc(__yt2) .. vloc(__yt1)}
\fmfi{fermion}{vloc(__xb1) .. vloc(__xb2)}
\end{fmfgraph*}}
\hspace{10mm} &  \text{\large$ 0 $}\hspace{2mm}
\\[12mm]
\hspace{-8mm}
\parbox{23mm}{
\begin{fmfgraph*}(23,13.)
\fmfstraight
\fmfleftn{l}{7}\fmfrightn{r}{7}\fmfbottomn{b}{7}\fmftopn{t}{7}
\fmf{phantom}{l2,vb1,mb1,vb2,r2}
\fmf{phantom}{l6,vt1,mt1,vt2,r6}
\fmf{phantom}{l2,xb1,mb1,xb2,r2}
\fmf{phantom}{l6,xt1,mt1,xt2,r6}
\fmf{phantom}{l2,yb1,mb1,yb2,r2}
\fmf{phantom}{l6,yt1,mt1,yt2,r6}
\fmf{phantom}{l4,c,r4}
\fmf{phantom}{l4,clb,r4}
\fmf{phantom}{l4,clt,r4}
\fmf{phantom}{l4,crb,r4}
\fmf{phantom}{l4,crt,r4}
\fmffreeze
\fmfshift{3 right}{xb1}
\fmfshift{3 down}{xb1}
\fmfshift{3 left}{xt2}
\fmfshift{3 down}{xt2}
\fmfshift{3 right}{xt1}
\fmfshift{3 down}{xt1}
\fmfshift{3 left}{xb2}
\fmfshift{3 down}{xb2}
\fmfshift{3 right}{yb1}
\fmfshift{3 up}{yb1}
\fmfshift{3 left}{yt2}
\fmfshift{3 up}{yt2}
\fmfshift{3 right}{yt1}
\fmfshift{3 up}{yt1}
\fmfshift{3 left}{yb2}
\fmfshift{3 up}{yb2}
\fmfshift{3 left}{clb}
\fmfshift{3 down}{clb}
\fmfshift{3 left}{clt}
\fmfshift{3 up}{clt}
\fmfshift{3 right}{crb}
\fmfshift{3 down}{crb}
\fmfshift{3 right}{crt}
\fmfshift{3 up}{crt}
\fmfv{d.sh=circle,d.f=empty,d.si=8,background=(.6235,,.7412,,1)}{vb1}
\fmfv{d.sh=circle,d.f=empty,d.si=8,background=(1,,.51,,.5)}{vb2}
\fmfv{d.sh=circle,d.f=empty,d.si=8,background=(.6235,,.7412,,1)}{vt1}
\fmfv{d.sh=circle,d.f=empty,d.si=8,background=(1,,.51,,.5)}{vt2}
\fmfv{d.sh=circle,d.f=empty,d.si=8,background=(1,,.51,,.5)}{c}
\fmf{dbl_plain}{l2,vb1}
\fmf{dbl_plain}{l6,vt1}
\fmf{dashes}{r2,vb2}
\fmf{dashes}{r6,vt2}
\fmfset{arrow_len}{2.2mm}
\fmfi{fermion}{ vloc(__crt).. vloc(__xt2)}
\fmfi{fermion}{vloc(__clt) ..vloc(__xt1)}
\fmfi{fermion}{vloc(__yb2)..vloc(__crb) }
\fmfi{fermion}{ vloc(__yb1)..vloc(__clb)}
\fmfi{fermion}{vloc(__yt2) .. vloc(__yt1)}
\fmfi{fermion}{vloc(__xb1) .. vloc(__xb2)}
\end{fmfgraph*}}
& 
\parbox{23mm}{
\begin{fmfgraph*}(23,13.)
\fmfstraight
\fmfleftn{l}{7}\fmfrightn{r}{7}\fmfbottomn{b}{7}\fmftopn{t}{7}
\fmf{phantom}{l2,vb1,mb1,vb2,r2}
\fmf{phantom}{l6,vt1,mt1,vt2,r6}
\fmf{phantom}{l2,xb1,mb1,xb2,r2}
\fmf{phantom}{l6,xt1,mt1,xt2,r6}
\fmf{phantom}{l2,yb1,mb1,yb2,r2}
\fmf{phantom}{l6,yt1,mt1,yt2,r6}
\fmf{phantom}{l4,c,r4}
\fmf{phantom}{l4,clb,r4}
\fmf{phantom}{l4,clt,r4}
\fmf{phantom}{l4,crb,r4}
\fmf{phantom}{l4,crt,r4}
\fmffreeze
\fmfshift{3 right}{xb1}
\fmfshift{3 down}{xb1}
\fmfshift{3 left}{xt2}
\fmfshift{3 down}{xt2}
\fmfshift{3 right}{xt1}
\fmfshift{3 down}{xt1}
\fmfshift{3 left}{xb2}
\fmfshift{3 down}{xb2}
\fmfshift{3 right}{yb1}
\fmfshift{3 up}{yb1}
\fmfshift{3 left}{yt2}
\fmfshift{3 up}{yt2}
\fmfshift{3 right}{yt1}
\fmfshift{3 up}{yt1}
\fmfshift{3 left}{yb2}
\fmfshift{3 up}{yb2}
\fmfshift{3 left}{clb}
\fmfshift{3 down}{clb}
\fmfshift{3 left}{clt}
\fmfshift{3 up}{clt}
\fmfshift{3 right}{crb}
\fmfshift{3 down}{crb}
\fmfshift{3 right}{crt}
\fmfshift{3 up}{crt}
\fmfv{d.sh=circle,d.f=empty,d.si=8,background=(.6235,,.7412,,1)}{vb1}
\fmfv{d.sh=circle,d.f=empty,d.si=8,background=(.6235,,.7412,,1)}{vb2}
\fmfv{d.sh=circle,d.f=empty,d.si=8,background=(.6235,,.7412,,1)}{vt1}
\fmfv{d.sh=circle,d.f=empty,d.si=8,background=(.6235,,.7412,,1)}{vt2}
\fmfv{d.sh=circle,d.f=empty,d.si=8,background=(1,,.51,,.5)}{c}
\fmf{dbl_plain}{l2,vb1}
\fmf{dbl_plain}{l6,vt1}
\fmf{dbl_plain}{r2,vb2}
\fmf{dbl_plain}{r6,vt2}
\fmfset{arrow_len}{2.2mm}
\fmfi{fermion}{vloc(__xt2) .. vloc(__crt)}
\fmfi{fermion}{ vloc(__clt)..vloc(__xt1)}
\fmfi{fermion}{vloc(__crb) ..vloc(__yb2)}
\fmfi{fermion}{vloc(__yb1) ..vloc(__clb)}
\fmfi{fermion}{vloc(__yt2) .. vloc(__yt1)}
\fmfi{fermion}{vloc(__xb1) .. vloc(__xb2)}
\end{fmfgraph*}} 
\hspace{10mm} &  \text{\large$ 0 $}
\hspace{2mm} \\[10mm]
\hspace{-8mm} 
 \text{\large$ 0 $} &  \hspace{-10mm} \text{\large$ 0 $} &  \text{\large$ 0 $}\hspace{2mm}
\end{pmatrix}
\end{align*}
\end{fmffile}   
\vspace{-3mm}

\caption{\fign{Feyn31}  The kernel matrix ${\cal V}^{(1)}$ defined as the rank-3 version of $V^{(1)}$, illustrated in \fig{Feyn1}.}
\end{center}
\end{figure}

\begin{figure}[t]
\begin{center}
\begin{fmffile}{Wmat}
\begin{align*}
{\cal W} \hspace{1mm} & = \hspace{2mm}
\begin{pmatrix}  \hspace{2mm}
 \text{\large$ 0 $} &  \hspace{8mm}
 \text{\large$ 0 $} &  \hspace{8mm}
\parbox{13mm}{
\begin{fmfgraph*}(13,13)
\fmfstraight
\fmfleftn{l}{7}\fmfrightn{r}{7}\fmfbottomn{b}{4}\fmftopn{t}{4}
\fmf{phantom}{r2,mb1,vb2,l2}
\fmf{phantom}{r6,mt1,vt2,l6}
\fmf{phantom}{r2,mb1,xb2,l2}
\fmf{phantom}{r6,mt1,xt2,l6}
\fmf{phantom}{r2,mb1,yb2,l2}
\fmf{phantom}{r6,mt1,yt2,l6}
\fmf{phantom,tension=3}{r2,mb1}
\fmf{phantom,tension=3}{r6,mt1}
\fmffreeze
\fmfshift{3 down}{r2}
\fmfshift{3 up}{r6}
\fmfshift{3 right}{xt2}
\fmfshift{3 down}{xt2}
\fmfshift{3 right}{xb2}
\fmfshift{3 down}{xb2}
\fmfshift{3 right}{yt2}
\fmfshift{3 up}{yt2}
\fmfshift{3 right}{yb2}
\fmfshift{3 up}{yb2}
\fmfv{d.sh=circle,d.f=empty,d.si=8,background=(1,,.51,,.5)}{vb2}
\fmfv{d.sh=circle,d.f=empty,d.si=8,background=(1,,.51,,.5)}{vt2}
\fmf{dashes}{l2,vb2}
\fmf{dashes}{l6,vt2}
\fmfset{arrow_len}{2.2mm}
\fmfi{fermion}{vloc(__xt2) {right}..tension 1 ..{left}vloc(__vb2)}
\fmfi{fermion}{vloc(__r6) .. vloc(__yt2)}
\fmfi{fermion}{vloc(__xb2) .. vloc(__r2)}
\end{fmfgraph*}}
\hspace{2mm}
\\[8mm]\hspace{2mm}
 \text{\large$ 0 $} &  \hspace{8mm} \text{\large$ 0 $}
& \hspace{8mm}
\parbox{13mm}{
\begin{fmfgraph*}(13,13)
\fmfstraight
\fmfleftn{l}{7}\fmfrightn{r}{7}\fmfbottomn{b}{4}\fmftopn{t}{4}
\fmf{phantom}{r2,mb1,vb2,l2}
\fmf{phantom}{r6,mt1,vt2,l6}
\fmf{phantom}{r2,mb1,xb2,l2}
\fmf{phantom}{r6,mt1,xt2,l6}
\fmf{phantom}{r2,mb1,yb2,l2}
\fmf{phantom}{r6,mt1,yt2,l6}
\fmf{phantom,tension=3}{r2,mb1}
\fmf{phantom,tension=3}{r6,mt1}
\fmffreeze
\fmfshift{3 down}{r2}
\fmfshift{3 up}{r6}
\fmfshift{3 right}{xt2}
\fmfshift{3 down}{xt2}
\fmfshift{3 right}{xb2}
\fmfshift{3 down}{xb2}
\fmfshift{3 right}{yt2}
\fmfshift{3 up}{yt2}
\fmfshift{3 right}{yb2}
\fmfshift{3 up}{yb2}
\fmfv{d.sh=circle,d.f=empty,d.si=8,background=(.6235,,.7412,,1)}{vb2}
\fmfv{d.sh=circle,d.f=empty,d.si=8,background=(.6235,,.7412,,1)}{vt2}
\fmf{dbl_plain}{l2,vb2}
\fmf{dbl_plain}{l6,vt2}
\fmfset{arrow_len}{2.2mm}
\fmfi{fermion}{vloc(__yb2) {right}..tension 1 ..{left}vloc(__xt2)}
\fmfi{fermion}{vloc(__r6) .. vloc(__yt2)}
\fmfi{fermion}{vloc(__xb2) .. vloc(__r2)}
\end{fmfgraph*}}
\hspace{2mm}
\\[8mm]\hspace{2mm}
\parbox{13mm}{
\begin{fmfgraph*}(13,13)
\fmfstraight
\fmfleftn{l}{7}\fmfrightn{r}{7}\fmfbottomn{b}{4}\fmftopn{t}{4}
\fmf{phantom}{l2,mb1,vb2,r2}
\fmf{phantom}{l6,mt1,vt2,r6}
\fmf{phantom}{l2,mb1,xb2,r2}
\fmf{phantom}{l6,mt1,xt2,r6}
\fmf{phantom}{l2,mb1,yb2,r2}
\fmf{phantom}{l6,mt1,yt2,r6}
\fmf{phantom,tension=3}{l2,mb1}
\fmf{phantom,tension=3}{l6,mt1}
\fmffreeze
\fmfshift{3 down}{l2}
\fmfshift{3 up}{l6}
\fmfshift{3 left}{xt2}
\fmfshift{3 down}{xt2}
\fmfshift{3 left}{xb2}
\fmfshift{3 down}{xb2}
\fmfshift{3 left}{yt2}
\fmfshift{3 up}{yt2}
\fmfshift{3 left}{yb2}
\fmfshift{3 up}{yb2}
\fmfv{d.sh=circle,d.f=empty,d.si=8,background=(1,,.51,,.5)}{vb2}
\fmfv{d.sh=circle,d.f=empty,d.si=8,background=(1,,.51,,.5)}{vt2}
\fmf{dashes}{r2,vb2}
\fmf{dashes}{r6,vt2}
\fmfset{arrow_len}{2.2mm}
\fmfi{fermion}{vloc(__yb2) {left}..tension 1 ..{right}vloc(__xt2)}
\fmfi{fermion}{vloc(__yt2) .. vloc(__l6)}
\fmfi{fermion}{vloc(__l2) .. vloc(__xb2)}
\end{fmfgraph*}}
&
\hspace{8mm}
\parbox{13mm}{
\begin{fmfgraph*}(13,13)
\fmfstraight
\fmfleftn{l}{7}\fmfrightn{r}{7}\fmfbottomn{b}{4}\fmftopn{t}{4}
\fmf{phantom}{l2,mb1,vb2,r2}
\fmf{phantom}{l6,mt1,vt2,r6}
\fmf{phantom}{l2,mb1,xb2,r2}
\fmf{phantom}{l6,mt1,xt2,r6}
\fmf{phantom}{l2,mb1,yb2,r2}
\fmf{phantom}{l6,mt1,yt2,r6}
\fmf{phantom,tension=3}{l2,mb1}
\fmf{phantom,tension=3}{l6,mt1}
\fmffreeze
\fmfshift{3 down}{l2}
\fmfshift{3 up}{l6}
\fmfshift{3 left}{xt2}
\fmfshift{3 down}{xt2}
\fmfshift{3 left}{xb2}
\fmfshift{3 down}{xb2}
\fmfshift{3 left}{yt2}
\fmfshift{3 up}{yt2}
\fmfshift{3 left}{yb2}
\fmfshift{3 up}{yb2}
\fmfv{d.sh=circle,d.f=empty,d.si=8,background=(.6235,,.7412,,1)}{vb2}
\fmfv{d.sh=circle,d.f=empty,d.si=8,background=(.6235,,.7412,,1)}{vt2}
\fmf{dbl_plain}{r2,vb2}
\fmf{dbl_plain}{r6,vt2}
\fmfset{arrow_len}{2.2mm}
\fmfi{fermion}{vloc(__xt2) {left}..tension 1 ..{right}vloc(__yb2)}
\fmfi{fermion}{vloc(__yt2) .. vloc(__l6)}
\fmfi{fermion}{vloc(__l2) .. vloc(__xb2)}
\end{fmfgraph*}}
&\hspace{7.5mm}
\parbox{15mm}{
\begin{fmfgraph*}(15,13)
\fmfstraight
\fmfleftn{l}{7}\fmfrightn{r}{7}\fmfbottomn{b}{5}\fmftopn{t}{5}
\fmf{phantom}{l1,ll1,m1,rr1,r1}
\fmf{phantom,tension=1.3}{ll1,m1,rr1}
\fmf{phantom}{l7,ll7,m7,rr7,r7}
\fmf{phantom,tension=1.3}{ll7,m7,rr7}
\fmf{phantom,tension=10}{m7,delta,m1}
\fmffreeze
\fmfshift{2.5 up}{l1}
\fmfshift{2.5 up}{r1}
\fmfshift{2.5 down}{l7}
\fmfshift{2.5 down}{r7}
\fmfset{arrow_len}{2.2mm}
\fmf{fermion}{ll7,l7}
\fmf{fermion}{l1,ll1}
\fmf{fermion}{rr1,r1}
\fmf{fermion}{r7,rr7}
\fmfv{d.sh=square,d.f=empty,d.si=14,background=(0.5216,,.7137,,.13333)}{delta}
\fmfiv{l=$\Delta$,l.a=180,l.d=.00w}{c}
\end{fmfgraph*}}
\end{pmatrix}
\end{align*}
\end{fmffile}   
\vspace{-3mm}

\caption{\fign{FeynW}  Elements of the kernel matrix ${\cal W}$: ${\cal W}_{31} = \bar N_M$ and ${\cal W}_{32} = \bar N_D$, are $q\q$-irreducible amplitudes for $MM\rightarrow q\q$ and $D\D \rightarrow q\q$, respectively; ${\cal W}_{13} = N_M$ and ${\cal W}_{23} =  N_D$, are the corresponding amplitudes for $q\q\rightarrow MM$ and $q\q\rightarrow D\D$.  Illustrated are models for these amplitudes as given by \eqs{Ns}. ${\cal W}_{33} = \Delta$ is the $q\q$ amplitude consisting of correction terms needed to make \eq{eq3} exact.}
\end{center}
\end{figure}

\section{Derivation}

The main results of this paper have been summarised above, and are embodied in the exact tetraquark equations of \eq{eq3}. In this section, we aim to provide as short a derivation of these equations as possible, but without sacrificing rigour or clarity. For this purpose we consider only the non-identical meson case, noting that the additional step of symmetrising $MM$ states for the case of identical mesons, can be implemented simply by referring to Ref.\  \cite{Kvinikhidze:2023djp} where this procedure is described.

To derive \eq{eq3},  we incorporate $q\q$ absorption into the unified tetraquark equations of Ref.\ \cite{Kvinikhidze:2023djp}, by exploiting the same procedure as we previously used in Ref.\ \cite{Kvinikhidze:2021kzu} to incorporate $q\q$ absorption into the tetraquark model of the Giessen group. This procedure is described in Sect.\ 2.1. In order to derive the precise mathematical expressions corresponding to the diagrams of Figs.\ 1 - 7, it is also necessary to discuss the derivation of the unified tetraquark equations themselves, at least for the case of non-identical mesons. Although this derivation has been previously given in a part of Ref.\ \cite{Kvinikhidze:2023djp}, for completeness we reproduce this derivation in Sect.\ 2.2 below, although using a somewhat simpler notation.

\subsection{Describing the tetraquark in the presence of $q\q$ annihilation}

{ In the absence of $q\q$ annihilation (either because annihilation is being neglected or because quantum numbers forbid it), the tetraquark can be simply defined as the four-body bound state of two quarks and two antiquarks as signalled by a pole in the full $2q2\q$ Green function $G^{(4)}$. However, in the presence of $q\q$ annihilation (when annihilation is physically realisable and taken into account), a pole in $G^{(4)}$ becomes an insufficient criterion for a tetraquark, as can be seen from the exact field-theoretic expression for $G^{(4)}$:
\be
G^{(4)} = G_{ir}^{(4)} + G_{ir}^{(4-2)} G_0^{(2)}{}^{-1} G^{(2)}G_0^{(2)}{}^{-1}G_{ir}^{(2-4)} ,  \eqn{exact}
\ee
where $G_{ir}^{(4)}$ is the $q\q$-irreducible part of $G^{(4)}$, $G_0^{(2)}$ is the disconnected part of the two-body $q\bar q$ Green function $G^{(2)}$ (corresponding to the independent propagation of $q$ and $\q$ in the $s$ channel), and $G_{ir}^{(2-4)}$ ($G_{ir}^{(4-2)}$) is the sum of all
$q\bar q$-irreducible diagrams corresponding to the transition $q\bar q\leftarrow 2q2\bar q$ ($2q2\bar q\leftarrow q\bar q$). Equation (\ref{exact}) shows that ${\it any}$ pole in the $q\q$ Green function $G^{(2)}$, even if not associated with a tetraquark,  is automatically a pole in $G^{(4)}$. 

Here we consider the case of the tetraquark in the presence of annihilation, meaning that the transition Green functions $G_{ir}^{(4-2)}$ and $G_{ir}^{(2-4)}$ in \eq{exact} are not zero. For such a case, both $G^{(4)}$ and $G^{(2)}$ will display simultaneous poles corresponding to a tetraquark of mass $M$, so that as $P^2\rightarrow M^2$ where $P$ is the total four-momentum of each system, 
\be
G^{(4)}\rightarrow i \frac{\Psi \bar\Psi}{P^2-M^2}, \hspace{1cm}
G^{(2)}\rightarrow i \frac{G_0^{(2)}\Gamma^* \bar\Gamma^*G_0^{(2)}}{P^2-M^2}.
\eqn{566}
\ee
In \eq{566}, $\Psi$ is the tetraquark  4-body $(2q2\q)$ bound state wave function, while $\Gamma^*$ is the form factor for the disintegration of a  tetraquark  into a $q\q$ pair.

As previously introduced in Ref.\ \cite{Kvinikhidze:2021kzu}, our key idea for describing a tetraquark in the presence on $q\q$ annihilation, is to: (i) express $G_{ir}^{(4)}$ in terms of some model for the $2q2\q$ system in absence of $q\q$ annihilation, and (ii) to then
express the $q\q$ kernel $K^{(2)}$, defined through the Dyson equation
\be
  G^{(2)} = G_0^{(2)}+G_0^{(2)}K^{(2)}G^{(2)},   \eqn{G2}
\ee
in terms of  $G_{ir}^{(4)}$ as
\be
K^{(2)} = \Delta + A^{(2-4)}G_{ir}^{(4)} A^{(4-2)}       \eqn{key}
\ee
where operators $A^{(4-2)}$ ($A^{(2-4)}$) are $q\q$-irreducibe amplitudes describing the transitions $2q 2\q\leftarrow q\q$ ($ q\q \leftarrow 2q2\q$) in the chosen model, and $\Delta$ is defined as the $q\q$-irreducible $q\q$ four-point function consisting of all contributions not accounted for by the last term of \eq{key}.  A feature of this approach is that no matter what model is chosen for $G_{ir}^{(4)}$ and $A^{(4-2)}$ ($A^{(2-4)}$), \eq{key} is exact in view of the definition of $\Delta$. Moreover, all terms contained in $\Delta$ are known thanks to its clear definition; for example, it contains sums of multiple-gluon exchanges. This fact enables one to systematically improve the approximations used in modelling the tetraquark, as for example, to take into account one-gluon exchange explicitly.

Furthermore, as the purpose of this work is to extend the unified tetraquark equations to include $q\q$ annihilation and other corrections, we assume that $G_{ir}^{(4)}$ (which is the Green function determining the unified tetraquark equations) has a pole at $P^2=M_0^2$, where $M_0$ is the tetraquark mass described by these equations; thus
\be
G_{ir}^{(4)} = i \frac{\Psi_0 \bar\Psi_0}{P^2-M_0^2} +R
\eqn{Gir_pole}
\ee
where $\Psi_0$ is the corresponding tetraquark wave function, and $R$ is a  background term. It is then straightforward to prove\footnote{It follows from the second of \eqs{566}, \eq{G2} and \eq{key} that
\[
\Gamma^* = K^{(2)} G_0^{(2)}\Gamma^* = \left[\Delta + A^{(2-4)}G_{ir}^{(4)} A^{(4-2)}\right] G_0^{(2)}\Gamma^*.
\]
Using \eq{Gir_pole} with $P^2=M^2$ in the above expression, and defining $X= \bar\Psi_0  A^{(4-2)} G_0^{(2)}\Gamma^*$, one obtains
\[
X =  \bar \Psi_0 A^{(4-2)}  \left[G_0^{(2)}{}^{-1}-\Delta- A^{(2-4)} R A^{(4-2)} \right]^{-1} A^{(2-4)} \Psi_0 \frac{i}{M^2-M_0^2} X
\]
which then implies \eq{MM0}.
}
\be
M^2 = M^2_0 + i \bar \Psi_0 A^{(4-2)} \left[G_0^{(2)}{}^{-1} -\Delta- A^{(2-4)} R A^{(4-2)} \right]^{-1} A^{(2-4)}\Psi_0 \eqn{MM0}
\ee
which shows that in the absence of annihilation and other corrections (described by $\Delta$), the tetraquark's mass $M_0$ plays the role of a ``bare" mass, and that the inclusion of annihilation and other corrections, then shifts the bare mass to the physical tetraquark mass $M$.

Here we apply this approach to the case where $A^{(2-4)}G_{ir}^{(4)} A^{(4-2)}$ is expressed in terms of the tetraquark model as described by the unified tetraquark equations,  \eq{eq}, and as derived  specifically in Sect.\ 2.2 below. In particular, we write this term as}
\be
A^{(2-4)}G_{ir}^{(4)} A^{(4-2)} = \bar N G N
\ee
where $G$ is the Green function in $MM-D\D$ space generated by the kernel of \eq{eq}, thus
\be
G = D + G V D, \hspace{1cm} V = V^{(0)} + V^{(1)} + \dots,
\ee
and $\bar N$ (simlarly $N$) is a matrix of transition amplitudes $\bar N_M$ (for $q\q \leftarrow MM$) and $\bar N_D$ (for $q\q \leftarrow D\D$).  We follow Ref.\ \cite{Kvinikhidze:2014yqa} where the simplest model for such amplitudes was realised, namely those described  graphically as
\be
\bar N = \begin{pmatrix} \bar N_M & \bar N_D \end{pmatrix}
= 
\begin{fmffile}{barN}
\begin{pmatrix}\hspace{.5mm}
\parbox{13mm}{
\begin{fmfgraph*}(13,13)
\fmfstraight
\fmfleftn{l}{7}\fmfrightn{r}{7}\fmfbottomn{b}{4}\fmftopn{t}{4}
\fmf{phantom}{l2,mb1,vb2,r2}
\fmf{phantom}{l6,mt1,vt2,r6}
\fmf{phantom}{l2,mb1,xb2,r2}
\fmf{phantom}{l6,mt1,xt2,r6}
\fmf{phantom}{l2,mb1,yb2,r2}
\fmf{phantom}{l6,mt1,yt2,r6}
\fmf{phantom,tension=3}{l2,mb1}
\fmf{phantom,tension=3}{l6,mt1}
\fmffreeze
\fmfshift{3 down}{l2}
\fmfshift{3 up}{l6}
\fmfshift{3 left}{xt2}
\fmfshift{3 down}{xt2}
\fmfshift{3 left}{xb2}
\fmfshift{3 down}{xb2}
\fmfshift{3 left}{yt2}
\fmfshift{3 up}{yt2}
\fmfshift{3 left}{yb2}
\fmfshift{3 up}{yb2}
\fmfv{d.sh=circle,d.f=empty,d.si=8,background=(1,,.51,,.5)}{vb2}
\fmfv{d.sh=circle,d.f=empty,d.si=8,background=(1,,.51,,.5)}{vt2}
\fmf{dashes}{r2,vb2}
\fmf{dashes}{r6,vt2}
\fmfset{arrow_len}{2.2mm}
\fmfi{fermion}{vloc(__yb2) {left}..tension 1 ..{right}vloc(__xt2)}
\fmfi{fermion}{vloc(__yt2) .. vloc(__l6)}
\fmfi{fermion}{vloc(__l2) .. vloc(__xb2)}
\end{fmfgraph*}}
&
\hspace{2mm}
\parbox{13mm}{
\begin{fmfgraph*}(13,13)
\fmfstraight
\fmfleftn{l}{7}\fmfrightn{r}{7}\fmfbottomn{b}{4}\fmftopn{t}{4}
\fmf{phantom}{l2,mb1,vb2,r2}
\fmf{phantom}{l6,mt1,vt2,r6}
\fmf{phantom}{l2,mb1,xb2,r2}
\fmf{phantom}{l6,mt1,xt2,r6}
\fmf{phantom}{l2,mb1,yb2,r2}
\fmf{phantom}{l6,mt1,yt2,r6}
\fmf{phantom,tension=3}{l2,mb1}
\fmf{phantom,tension=3}{l6,mt1}
\fmffreeze
\fmfshift{3 down}{l2}
\fmfshift{3 up}{l6}
\fmfshift{3 left}{xt2}
\fmfshift{3 down}{xt2}
\fmfshift{3 left}{xb2}
\fmfshift{3 down}{xb2}
\fmfshift{3 left}{yt2}
\fmfshift{3 up}{yt2}
\fmfshift{3 left}{yb2}
\fmfshift{3 up}{yb2}
\fmfv{d.sh=circle,d.f=empty,d.si=8,background=(.6235,,.7412,,1)}{vb2}
\fmfv{d.sh=circle,d.f=empty,d.si=8,background=(.6235,,.7412,,1)}{vt2}
\fmf{dbl_plain}{r2,vb2}
\fmf{dbl_plain}{r6,vt2}
\fmfset{arrow_len}{2.2mm}
\fmfi{fermion}{vloc(__xt2) {left}..tension 1 ..{right}vloc(__yb2)}
\fmfi{fermion}{vloc(__yt2) .. vloc(__l6)}
\fmfi{fermion}{vloc(__l2) .. vloc(__xb2)}
\end{fmfgraph*}}\hspace{.5mm}
\end{pmatrix}
\end{fmffile}, \hspace{4mm} N = \begin{pmatrix} N_M \\[2mm] N_D \end{pmatrix} =
\begin{fmffile}{N}
\begin{pmatrix}\hspace{.5mm}
\parbox{13mm}{
\begin{fmfgraph*}(13,13)
\fmfstraight
\fmfleftn{l}{7}\fmfrightn{r}{7}\fmfbottomn{b}{4}\fmftopn{t}{4}
\fmf{phantom}{r2,mb1,vb2,l2}
\fmf{phantom}{r6,mt1,vt2,l6}
\fmf{phantom}{r2,mb1,xb2,l2}
\fmf{phantom}{r6,mt1,xt2,l6}
\fmf{phantom}{r2,mb1,yb2,l2}
\fmf{phantom}{r6,mt1,yt2,l6}
\fmf{phantom,tension=3}{r2,mb1}
\fmf{phantom,tension=3}{r6,mt1}
\fmffreeze
\fmfshift{3 down}{r2}
\fmfshift{3 up}{r6}
\fmfshift{3 right}{xt2}
\fmfshift{3 down}{xt2}
\fmfshift{3 right}{xb2}
\fmfshift{3 down}{xb2}
\fmfshift{3 right}{yt2}
\fmfshift{3 up}{yt2}
\fmfshift{3 right}{yb2}
\fmfshift{3 up}{yb2}
\fmfv{d.sh=circle,d.f=empty,d.si=8,background=(1,,.51,,.5)}{vb2}
\fmfv{d.sh=circle,d.f=empty,d.si=8,background=(1,,.51,,.5)}{vt2}
\fmf{dashes}{l2,vb2}
\fmf{dashes}{l6,vt2}
\fmfset{arrow_len}{2.2mm}
\fmfi{fermion}{vloc(__xt2) {right}..tension 1 ..{left}vloc(__vb2)}
\fmfi{fermion}{vloc(__r6) .. vloc(__yt2)}
\fmfi{fermion}{vloc(__xb2) .. vloc(__r2)}
\end{fmfgraph*}}\hspace{.5mm}
\\[6mm]
\parbox{13mm}{
\begin{fmfgraph*}(13,13)
\fmfstraight
\fmfleftn{l}{7}\fmfrightn{r}{7}\fmfbottomn{b}{4}\fmftopn{t}{4}
\fmf{phantom}{r2,mb1,vb2,l2}
\fmf{phantom}{r6,mt1,vt2,l6}
\fmf{phantom}{r2,mb1,xb2,l2}
\fmf{phantom}{r6,mt1,xt2,l6}
\fmf{phantom}{r2,mb1,yb2,l2}
\fmf{phantom}{r6,mt1,yt2,l6}
\fmf{phantom,tension=3}{r2,mb1}
\fmf{phantom,tension=3}{r6,mt1}
\fmffreeze
\fmfshift{3 down}{r2}
\fmfshift{3 up}{r6}
\fmfshift{3 right}{xt2}
\fmfshift{3 down}{xt2}
\fmfshift{3 right}{xb2}
\fmfshift{3 down}{xb2}
\fmfshift{3 right}{yt2}
\fmfshift{3 up}{yt2}
\fmfshift{3 right}{yb2}
\fmfshift{3 up}{yb2}
\fmfv{d.sh=circle,d.f=empty,d.si=8,background=(.6235,,.7412,,1)}{vb2}
\fmfv{d.sh=circle,d.f=empty,d.si=8,background=(.6235,,.7412,,1)}{vt2}
\fmf{dbl_plain}{l2,vb2}
\fmf{dbl_plain}{l6,vt2}
\fmfset{arrow_len}{2.2mm}
\fmfi{fermion}{vloc(__yb2) {right}..tension 1 ..{left}vloc(__xt2)}
\fmfi{fermion}{vloc(__r6) .. vloc(__yt2)}
\fmfi{fermion}{vloc(__xb2) .. vloc(__r2)}
\end{fmfgraph*}}\hspace{.5mm}
\end{pmatrix}
\end{fmffile} ,
\ee
and given analytically in the notation of Sect.\ 2.2.3 below, as
\begin{subequations}  \eqn{Ns}
\begin{alignat}{2} 
\bar N_M& =S_{23}\Gamma_{13} \Gamma_{24},&\hspace{5mm}
\bar N_D &=S_{23}\Gamma_{12} \Gamma_{34} , \eqn{bN} \\[2mm]
N_M&= \bar\Gamma_{13} \bar\Gamma_{24},&\hspace{5mm}
N_D &= \bar\Gamma_{12}^p \bar\Gamma_{34} S_{23} \eqn{N},
\end{alignat}
\end{subequations}
where $S_{23}$ is the quark propagator connecting quark lines 2 and 3.  Expressions for the kernels making up $V$ are also derived in Sect.\ 2.2.3, with $V^{(0)}$ being given by \eq{V0mat}, and $V^{(1)}$ by \eq{V1mat}.
Thus
\be
K^{(2)} = \Delta + \bar N \left[D(1- VD)^{-1}) \right] N  .  \eqn{K2}
\ee
It is also evident from \eq{G2} and the second of the relations in \eq{566}, that the tetraquark state will also satisfy the two-body equation
\be
\Gamma^*=K^{(2)}G_0^{(2)}\Gamma^* .   \eqn{2b-tetr}
\ee
Using \eq{K2} in \eq{2b-tetr}, one obtains
\be
\Gamma^*=\Delta G_0^{(2)}\Gamma^* + \bar N D \phi  \eqn{Gamma}
\ee
where
\be
\phi = V D \phi + N G_0^{(2)} \Gamma^*,    \eqn{chi}
\ee
with $\phi$ being the column of tetraquark form factors $\phi_M$ and $\phi_D$ as in \eq{phiD}. By introducing a column of tetraquark form factors that also includes the form factor $\Gamma^*$,
\be
\varphi = \begin{pmatrix} \phi_M \\ \phi_D \\\Gamma^*\end{pmatrix},
\ee
and defining $G_0^{(2)} \equiv Q\bar Q$, \eq{Gamma} and \eq{chi} can be expressed succinctly as the rank-3 matrix equation \eq{eq3}.

\subsection{Unified tetraquark equations}

{The exact unified tratraquark equations, \eq{eq3}, take as an input the kernel of the unified tetraquark equations of  Ref.\ \cite{Kvinikhidze:2023djp}, including the separate form factors for the $M \leftrightarrow q\q$,  $D \leftrightarrow q q$, and $\D \leftrightarrow \q \q$ processes that may be used to model the input $MM \leftrightarrow q\q$ and  $D\D \leftrightarrow q\q$ transition amplitudes, as for example in \eqs{Ns}. For this reason, it is essential to refer to the derivation of the unified tetraquark equations where all the expressions used in the current work as input, are developed.
Although this derivation can be found in full in Ref.\ \cite{Kvinikhidze:2023djp}, for the purposes of the current work, it is sufficient to consider this deriviation just for the case of non-identical mesons. For completeness, we present this derivation here, following closely the presentation of Ref.\ \cite{Kvinikhidze:2023djp}, but using a somewhat simpler notation. }

\subsubsection{Four-body equations for distinguishable quarks}

We consider the 4-body system consisting of $2q$ and $2\q$ treated as distinguishable particles where $q\q$ annihilation (or creation) is not allowed. The 4-body Green function describing this system is then $G^{(4)}_{ir}$, the $q\q$-irreducible part of the full Green function $G^{(4)}$, as in  \eq{exact}.  One can then introduce the 4-body  kernel $K$ and corresponding t matrix $T$ (both $q\q$-irreducible) through the equations
\begin{align}
G^{(4)}_{ir}  &= G_0 + G_0 K  G^{(4)}_{ir},\hspace{1cm} G^{(4)}_{ir} = G_0 + G_0 T  G_0
\end{align}
where $G_0$ describes the free propagation of all four particles.  We assign labels 1,2 to the quarks and 3,4 to the antiquarks and following Ref.\ \cite{Khvedelidze:1991qb}, introduce an index $a \in \left\{12,13,14,23,24,34\right\}$ which enumerates the six possible pairs of particles. Similarly, we introduce the double index $aa'  \in \left\{(13,24), (14,23),(12,34)\right\}$ which enumerates the three possible two-pairs of particles, and use the Greek  index $\A$ as an abbreviation for $aa'$ such that $\A=1$ denotes  $aa'=(13,24)$,  $\A=2$ denotes $aa'=(14,23)$, and  $\A=3$ denotes $aa'=(12,34)$.

Using this labelling scheme, it is useful to consider the Green function $G_{aa'}$ defined as the part of $G^{(4)}_{ir}$ where all interactions are switched off except those within the pairs $a$ and $a'$. One can then introduce the corresponding kernel $K_{aa'}$ and t matrix $T_{aa'}$ through the equations
\begin{align}
G_{aa'} = G_0 + G_0 K_{aa'} G_{aa'}, \hspace{1cm} G_{aa'} = G_0 + G_0 T_{aa'} G_0.
\end{align}
Note that we drop the superscript indicating the number of particles described, when this is otherwise implied by the meaning of the subscripts.
It is evident that in the theory where only pairwise interactions are allowed, one has that
\be
K =\sum_{aa'} K_{aa'} =\sum_{\A} K_{\A} . \eqn{pair}
\ee
This is a key expression for $K$ as it is of similar form to that describing the kernel of a 3-body system interacting via pairwise interactions, thereby leading to the formulation of 4-body equations by analogy to those of 3-body equations \cite{Khvedelidze:1991qb}.
It is also clear that
\be
G_{aa'} = G_a G_a'
\ee
where $G_a$   (similarly $G_{a'}$) is the full 2-body Green function for the scattering of particles of pair $a$, with corresponding 2-body kernel $K_{a}$  and t matrix $T_a$  defined through
\be
G_{a} = G_a^{0} + G_a^{0} K_{a} G_{a},\hspace{1cm} G_{a} = G_a^{0} + G_a^{0} T_{a} G_a^{0},
\ee
where $G_a^0$ describes the free propagation of the particles in pair $a$. Using the above equations it is then easy to show \eq{Taa} and similarly that
\be
K_{aa'}=K_a G_{a'}^{0}{}^{-1}+K_{a'}G_{a}^{0}{}^{-1} -K_a K_{a'}, \eqn{Kaa}
\ee
where the presence of a minus sign in the last term of \eq{Kaa} is necessary to avoid overcounting.
 
The $2q2\q$  bound state form factor for distinguishable quarks is then
\be
 \Phi = K G_0  \Phi.       \eqn{Phi}
 \ee
The four-body kernels $K_\A$ can be used to define the Faddeev components of $\Phi$ as
\be
\Phi_\A = K_\A G_0 \Phi,
\ee
so that
\be
\sum_\A \Phi_\A = \Phi.
\ee
From \eq{Phi} follow Faddeev-like equations for the components,
\be
\Phi_\alpha=T_\alpha\sum_\beta\bar\delta_{\alpha\beta}G_0 \Phi_\beta    \eqn{PhiFad}
\ee
where $\bar\delta_{\alpha\beta}=1-\delta_{\alpha\beta}$.

\subsubsection{Four-body equations for indistinguishable quarks}

The bound state equation for  two identical quarks $1,2$ and two identical antiquarks $3,4$, is modified from \eq{Phi} to
\be
\Phi=\frac{1}{4}K G_0\Phi ,  \eqn{ident2q2q}
\ee
where now the kernel $K$ is antisymmetric with respect to swapping quark or antiquark quantum numbers either in the initial or in the final state; that is,
 \be
 {\cal P}_{34}K={\cal P}_{12}K=K{\cal P}_{34}=K{\cal P}_{12}=-K   \eqn{Kas}
 \ee
 where the exchange operator $\PP_{ij}$ swaps the quantum numbers associated with particles $i$ and $j$ in the quantity on which it is operating.
The factor $ \frac{1}{4}$ in   \eq{ident2q2q} is a product of the combinatorial factors $\frac{1}{2}$, one for identical quarks and another for  identical antiquarks.

In order to distinguish the quantities $\Phi$, $K$, and $T$ in the present case of indistinguishable quarks, from their distinguishable-quark counterparts discussed in the previous section, we shall simply use a superscript $d$ to denote the quantities that apply to the case of distinguishable quarks.
Thus \eq{Phi} for distinguishable quarks shall now be written as
\be
 \Phi^d = K^d G_0  \Phi^d,    \eqn{dist2q2q}
 \ee
\eq{PhiFad} as
 \be
\Phi^d_\alpha=T^d_\alpha\sum_\beta\bar\delta_{\alpha\beta}G_0 \Phi^d_\beta ,   \eqn{PhidFad}
\ee
and \eq{Taa} as
\be
T^d_\A = T^d_a G_{a'}^{0}{}^{-1}+T^d_{a'}G_{a}^{0}{}^{-1}  + T^d_a T^d_{a'}.  \eqn{Taad}
\ee

The kernel $K$ that is antisymmetric in the way specified by \eq{Kas}, can be represented as
 \be
 K=(1-{\cal P}_{12})(1-{\cal P}_{34})K^d
 \ee
 where $K^d$ is symmetric with respect to swapping either quark or antiquark quantum numbers in the initial and final states simultaneously, ${\cal P}_{12}K^d{\cal P}_{12}={\cal P}_{34}K^d{\cal P}_{34}=K^d$. This symmetry property of $K^d$ can be written in the form of commutation relations
 \be
 [{\cal P}_{34},K^d]=[{\cal P}_{12},K^d]=0,    \eqn{Kdcom}
 \ee
and follows directly from  the following relations implied by \eqs{Kaa}:
\begin{subequations}\eqn{Ksym}
\begin{align}  
{\cal P}_{12}K^{d}_3{\cal P}_{12}& ={\cal P}_{34}K^{d}_3{\cal P}_{34}=K^{d}_3, \\
{\cal P}_{12}K^{d}_1{\cal P}_{12}& ={\cal P}_{34}K^{d}_1{\cal P}_{34}=K^{d}_2.
\end{align}
\end{subequations}
Due to the antisymmetry properties of $K$ as specified in \eq{Kas}, the solution of the identical particle bound state equation, \eq{ident2q2q}, is correspondingly antisymmetric; namely,  ${\cal P}_{34}\Phi={\cal P}_{12}\Phi=-\Phi$.
However, because $K^d$ usually corresponds to a fewer number of diagrams than $K$, rather than solving \eq{ident2q2q}, it may be more convenient to determine $\Phi$ by antisymmetrising the solution $\Phi^d$ of the bound state equation for distinguishable quarks, as
\be
\Phi=(1-{\cal P}_{12})(1-{\cal P}_{34})\Phi^d. \eqn{PhiAS}
\ee
Then, in view of the commutation relations of \eq{Kdcom}, if the solution $\Phi^d$
 exists, its antisymmetrized version as given by \eq{PhiAS}, also satisfies the bound state equation for distinguishable quarks, \eq{dist2q2q}, as well as the one for indistinguishable ones,  \eq{ident2q2q}:
\begin{align}
\Phi&=(1-{\cal P}_{12})(1-{\cal P}_{34})K^dG_0\Phi ^d \nn
& =K^dG_0(1-{\cal P}_{12})(1-{\cal P}_{34})\Phi ^d =K^dG_0\Phi  \nn
&= \frac{1}{4}K^dG_0(1-{\cal P}_{12})(1-{\cal P}_{34})\Phi 
=\frac{1}{4}K G_0\Phi.  \eqn{11**}
\end{align}
As $\Phi$ satisfies the same bound state equation as $\Phi^d$,
\be
\Phi= K^d G_0 \Phi ,  \eqn{PhiBS}
\ee
the kernels $K^{d}_\A$ can again be used to define Faddeev components, but this time for $\Phi$:
\be
\Phi_\A = K^{d}_\A G_0 \Phi,
\ee
where
\be
\sum_\A \Phi_\A = \Phi.
\ee
In view of \eqs{Ksym}, the Faddeev components $\Phi_\alpha$  have the following properties:
\begin{subequations} \eqn{P12Phi1= P34Phi1}\eqn{Phisym}
\begin{alignat}{2}
{\cal P}_{12}\Phi_3 & =-\Phi_3, & \hspace{1cm}
{\cal P}_{12}\Phi_1 & =-\Phi_2, \eqn{Phisyma}\\
{\cal P}_{34}\Phi_3 & =-\Phi_3, & \hspace{1cm}
{\cal P}_{34}\Phi_1 &=-\Phi_2.   \eqn{Phisymb}
\end{alignat}
\end{subequations}

Since $\Phi$ satisfies the same bound state equation as $\Phi^d$,  the components $\Phi_\A$ satisfy the same Faddeev-like equations as for distinguishable quarks, \eq{PhidFad}, 
 \be
\Phi_\alpha=T^d_\alpha\sum_\beta\bar\delta_{\alpha\beta}G_0 \Phi_\beta ,    \eqn{distj}
\ee
Without loss of generality,  we can assume that the solution of \eq{distj} has the symmetry properties of \eq{P12Phi1= P34Phi1} \cite{Kvinikhidze:2023djp}. We also note that the input 2-body t matrices $T^{d}_{12}$ and $T^{d}_{34}$ can be antisymmetrized by defining
\be
T_{12} = \frac{1}{2}(1-\PP_{12}) T^{d}_{12}, \hspace{5mm}
T_{34} = \frac{1}{2}(1-\PP_{34}) T^{d}_{34},
\ee
so that
\begin{subequations}  \eqn{Tanti}
\begin{align}
T_{12} \PP_{12} &= \PP_{12} T_{12} = - T_{12} ,\\
T_{34} \PP_{34} &= \PP_{34} T_{34} = - T_{34}.
\end{align}
\end{subequations}
By defining further that
\begin{subequations} 
\begin{align}
T_{13} & = T^{d}_{13}, \hspace{5mm} T_{24} = T^{d}_{24}, \hspace{5mm}
T_{14}  = T^{d}_{14}, \hspace{5mm} T_{23} = T^{d}_{23}, \\
T_{1} & = T^{d}_{1}, \hspace{5mm} T_{2} = T^{d}_{2}, \\
T_3& =T_{12} G_{12}^0{}^{-1}+T_{34}G_{34}^0{}^{-1}+T_{12}T_{34} , \eqn{T3=T12c}
\end{align}
\end{subequations}
we have made \eq{Taa} apply, with unchanged notation, also for the case of indistinguishable quarks. Moreover, one can now write \eq{distj} simply as
 \be
\Phi_\alpha=T_\alpha\sum_\beta\bar\delta_{\alpha\beta}G_0 \Phi_\beta ,    \eqn{distii}
\ee
where dropping the superscript $d$ from $T_\A^d$ for the case $\A=3$ can be justified by multiplying \eq{distj} for this case by $(1-\PP_{12})$ and using the symmetry properties of \eq{P12Phi1= P34Phi1}:
\begin{align}
\Phi_3 &=\frac{1}{2}(1-\PP_{12}) T^{d}_3 \frac{1}{2}(1-\PP_{34}) (\Phi_1+\Phi_2)\nn
&=T_3 (\Phi_1+\Phi_2).
\end{align}
The advantage of using $T_3$ rather than $T^d_3$ in this equation stems from the fact that the physical (antisymmetric) t matrices for $qq$ and $\q\q$ scattering are $T_{qq} = (1-{\cal P}_{12})T^d_{12} =2T_{12}$ and $T_{\q\q} = (1-{\cal P}_{34})T^d_{34} =2T_{34}$, respectively, therefore it is convenient to use the antisymmetric $T_{12}$ and $T_{34}$ as the input $qq$ and $\q\q$ t matrices. 

Writing out  \eqs{distii} in full, but to save notation without explicitly showing $G_0$'s, one has 
\begin{subequations}  \eqn{disti}
\begin{align}
\Phi_1&=T_1 (\Phi_2+\Phi_3), \eqn{distia}\\
\Phi_2&=T_2 (\Phi_3+\Phi_1) ,\eqn{distib}\\
\Phi_3&=T_3 (\Phi_1+\Phi_2)   . \eqn{distic}
\end{align}
\end{subequations}
For physical (antisymmetric) solutions of  \eqs{disti}, only two of these three equations are independent. For example,  \eq{distib} can be written as
\begin{align}
-{\cal P}_{12}\Phi_1&={\cal P}_{12}T_1{\cal P}_{12} (\Phi_3+\Phi_1) \nn
&={\cal P}_{12}T_1 (-\Phi_3 - \Phi_2)
\end{align}
where \eq{P12Phi1= P34Phi1} and $T_2={\cal P}_{12}T_1{\cal P}_{12}$ have been used; then, after a further application of ${\cal P}_{12}$, one obtains  \eq{distia}. Choosing \eq{distia} and \eq{distic} as the two independent equations, we can use $ \Phi_2 = -{\cal P}_{12}\Phi_1$ to obtain closed equations
\begin{subequations}  \eqn{1,3}
\begin{align}
\Phi_1&=T_1 (-{\cal P}_{12}\Phi_1+\Phi_3), \eqn{1,3a}\\
\Phi_3&=T_3 (\Phi_1-{\cal P}_{12}\Phi_1), \eqn{1,3b}
\end{align}
\end{subequations}
where, necessarily, ${\cal P}_{12}\Phi_3  = -\Phi_3$. Again, without loss of generality,  we can assume that the solutions of \eq{1,3} have the symmetry properties of \eq{P12Phi1= P34Phi1} \cite{Kvinikhidze:2023djp}. 
Equation (\ref{1,3b}) can be further simplified using ${\cal P}_{12}\Phi_1={\cal P}_{34}\Phi_1$  and the assumption that $T_{12}$ and $T_{34}$ are antisymmetric in their labels,  so that
\begin{align}
T_3 {\cal P}_{12}\Phi_1 &= (T_{12}+T_{34}+T_{12}T_{34}){\cal P}_{12}\Phi_1 =-T_3 \Phi_1.
\end{align}
In this way \eqs{1,3} take the form
\begin{subequations}\eqn{1,3****}
\begin{align}
\Phi_1&=T_1 (-{\cal P}_{12}\Phi_1+\Phi_3) \eqn{1,3a****} \\
\Phi_3&=2T_3 \Phi_1. \eqn{1,3b****}
\end{align}
\end{subequations}
Again, without loss of generality,  we choose a solution of \eqs{1,3****} which has all the symmetry properties of \eq{P12Phi1= P34Phi1}.

\subsubsection{Tetraquark equations with exposed  ${q\bar q (T_{q\q})}$, ${q q (T_{\q\q})}$, and ${\q\q (T_{qq})}$ 
channels}\label{dd-int}

Choosing \eqs{1,3****} as the four-body equations describing a tetraquark, they may be expressed in matrix form as
\be
\PPhi = \TT\RR\PPhi       \eqn{P,12}
\ee
where
{
\be
\PPhi = \begin{pmatrix} \Phi_1 \\  \Phi_3 \end{pmatrix} ,\hspace{2mm}
\TT = \begin{pmatrix} \frac{1}{2} T_1  & 0\\  0&T_3 \end{pmatrix}  ,\hspace{2mm}
\RR =    2\begin{pmatrix} -  {\cal P}_{12} & 1 \\  1 &0 \end{pmatrix} .
\ee}
Writing
\be
T_1 =T^\times_1+T^+_1 ,\hspace{1cm}
T_3 =T^\times_3+T^+_3 ,
\ee
where
\begin{subequations} \eqn{Tx}
\begin{alignat}{2}
T^\times_1&=T_{13}T_{24}, & \hspace{1cm} T^+_1&=T_{13}+T_{24}, \eqn{Txa}\\
T^\times_3& =T_{12}T_{34} & \hspace{1cm} T^+_3&=T_{12}+T_{34}, \eqn{Txb}
\end{alignat}
\end{subequations}
we have that
\be
\TT = \TT^\times + \TT^+
\ee
where
{
\be
\TT^\times= \begin{pmatrix} \frac{1}{2} T^\times_1  & 0\\  0&T^\times_3 \end{pmatrix}  ,\hspace{2mm}
\TT^+ = \begin{pmatrix} \frac{1}{2}T^+_1  & 0\\  0&T^+_3 \end{pmatrix}  .
\ee}
Thus
\be
\PPhi = (\TT^\times + \TT^+) \RR\PPhi 
\ee
and consequently
\be
\PPhi = (1- \TT^+\RR)^{-1} \TT^\times \RR\PPhi  .  \eqn{(-1)}
\ee

To be close to previous publications we choose a separable approximation for the two-body t matrices in $T^\times_1$ and $T^\times_3$ (but not necessarily in $T^+_1$ and $T^+_3$); namely, for $a \in \left\{13, 24,12, 34\right\}$ we take
\be
T_{a}=i\Gamma_{a} D_{a} \bar\Gamma_{a},
\ee
where $D_{a}=D_{a}(P_{a})$ is a propagator whose structure can be chosen to best describe the two-body t matrix $T_{a}$, and $\Gamma_{a}$ is a corresponding vertex function.  In the simplest case, one can follow previous publications and choose the pole approximation  where $D_{a}(P_{a})=1/(P_{a}^2-m_{a}^2)$ is the propagator for the bound particle (diquark, antidiquark, or meson) of mass $m_a$. In view of \eq{Tanti}, note that
\begin{subequations}
\begin{alignat}{2}
\PP_{12} \Gamma_{12} &= - \Gamma_{12}, & \hspace{1cm}  \bar\Gamma_{12} \PP_{12}&= - \bar\Gamma_{12}, \\
\PP_{34} \Gamma_{34} &= - \Gamma_{34}, & \hspace{1cm}  \bar\Gamma_{34} \PP_{34}&= - \bar\Gamma_{34}.
\end{alignat}
\end{subequations}
We can thus write
\be
\TT^\times = -\GG D\bGG     \eqn{Txsep}
\ee
where
{
\be
\GG= \begin{pmatrix} \Gamma_{13}\Gamma_{24}  & 0\\  0& \Gamma_{12}\Gamma_{34} \end{pmatrix}  ,\hspace{2mm}
D = \begin{pmatrix}  \frac{1}{2}D_{13}D_{24}  & 0\\  0& D_{12} D_{34} \end{pmatrix}  ,\hspace{2mm}
\bGG= \begin{pmatrix} \bar \Gamma_{13}\bar\Gamma_{24}  & 0\\  0& \bar\Gamma_{12}\bar\Gamma_{34} \end{pmatrix} .
\ee}
In this way $\TT^\times$ exposes intermediate state meson-meson $(D_{13}D_{24})$ and diquark-antidiquark $(D_{12}D_{34})$ channels. Using \eq{Txsep} in \eq{(-1)}, 
\be
\phi =-\bGG \RR (1- \TT^+\RR)^{-1} \GG D\phi  \eqn{2-step-eq}
\ee
where
\be
\phi = \bGG \RR\PPhi .  \eqn{2-step-phi}
\ee
In this way we obtain the bound state equation for $\phi$ in meson-meson $(MM)$ and diquark-antidiquark $(D\bar D)$  space,
\be 
\phi=V D\phi    \eqn{400-}
\ee
where the $2\times 2$ matrix potential (with reinserted $\Gf_0 $) is
\be
V =-\bGG \RR\Gf_0 (1- \TT^+\RR\Gf_0)^{-1} \GG.    \eqn{Vfull}
\ee
Expanding the term in square brackets in powers of  $\TT^+$ (i.e., with respect to the contribution of intermediate states $q\bar q (T_{q\q})$, $q q (T_{\q\q})$, and $\q\q (T_{qq})$),
\be
V = -\bGG  \RR\Gf_0 \left[1+ \TT^+\RR\Gf_0+\dots\right]\GG,   \eqn{V}
\ee
it turns out that each of the first two terms of this expansion corresponds to different existing approaches to modelling tetraquarks in terms of $MM-D\bar D$ coupled channels. In particular, the lowest order term
\begin{align}
V^{(0)} &=  -\bGG \RR \Gf_0 \GG  \nn
&=- 2\begin{pmatrix} \bar\Gamma_1 & 0\\  0&\bar\Gamma_3 \end{pmatrix}
\begin{pmatrix} -  {\cal P}_{12} & 1 \\  1 &0 \end{pmatrix} \Gf_0 
  \begin{pmatrix}\Gamma_1  & 0\\  0&\Gamma_3 \end{pmatrix}\nn[2mm]
 &= - 2\begin{pmatrix} -  \bGamma_1\PP_{12} \Gamma_1& \bar\Gamma_1\Gamma_3\\  
  \bar\Gamma_3  \Gamma_1&0 \end{pmatrix}, \eqn{V0mat}
\end{align}
where
\begin{subequations}
\begin{alignat}{2}
\bGamma_1  &= \bGamma_{13}\bGamma_{24}, & \hspace{1cm}  \Gamma_1  &= \Gamma_{13}\Gamma_{24}, \\
\bGamma_3  &= \bGamma_{12}\bGamma_{34}, & \hspace{1cm}  \Gamma_3  &= \Gamma_{12}\Gamma_{34},
\end{alignat}
\end{subequations}
consists of Feynman diagrams illustrated in \fig{Feyn0}, and corresponds to the Giessen  group (GG) model of Heupel {\it et al.} \cite{Heupel:2012ua} where tetraquarks are modelled by solving \eq{phi_Giessen}.

Similarly,  the first order correction (without the lowest order term included) is
\begin{align}
V^{(1)} &= -\bGG \RR  \Gf_0 \TT^+\RR\Gf_0\GG \nn
&=- 4\begin{pmatrix} \bar\Gamma_1 & 0\\  0&\bar\Gamma_3 \end{pmatrix}  \Gf_0
\begin{pmatrix} -  {\cal P}_{12} & 1 \\  1 &0 \end{pmatrix}  \begin{pmatrix} \frac{1}{2} T^+_1  & 0\\  0&T^+_3 \end{pmatrix} 
   \begin{pmatrix} -  {\cal P}_{12} & 1 \\  1 &0 \end{pmatrix} \Gf_0  
  \begin{pmatrix}\Gamma_1  & 0\\  0&\Gamma_3 \end{pmatrix} 
  \nn[2mm]
  &=-  2\begin{pmatrix} \bar\Gamma_1[{\cal P}_{12}  T^+_1 {\cal P}_{12}+2T^+_3]\Gamma_1 &-\bar\Gamma_1{\cal P}_{12}  T^+_1\Gamma_3 \\ 
   -\bar\Gamma_3T^+_1{\cal P}_{12}\Gamma_1 &2\bar\Gamma_3T^+_1 \Gamma_3\end{pmatrix} ,  \eqn{V1mat}
\end{align}
which consists of Feynman diagrams illustrated in \fig{Feyn1}, and corresponds to the Moscow group (MG) model of Faustov {\it et al.} \cite{Faustov:2020qfm} where they modelled tetraquarks by solving \eq{phi_Moscow}
albeit, with only diquark-antidiquark channels retained. It is an essential feature of this formulation, that it is the sum of the potentials $V^{(0)}$ and $V^{(1)}$, each associated with the separate approaches of the Giessen and Moscow groups, with tetraquarks modelled by the bound state equation
\be
\phi = [V^{(0)}+V^{(1)}] D \phi,  \eqn{VG+VF}
\ee
that results in a complete $MM-D\bar D$ coupled channel description up to first order in  $\TT^+$ [i.e., up to first order in intermediate states where one $2q$ pair ($qq$, $q\q$, or $\q\q$) is mutually interacting while the other $2q$ pair is spectating].

Finally we need to point out that for the case of identical mesons, the $MM$ states in the above $MM-D\bar D$ coupled channel description need to be symmetrised. This does not change the structure of any of the above equations, or the essential results. The details of this symmetrisation are given in Ref.\ \cite{Kvinikhidze:2023djp}.

\begin{acknowledgments}

A.N.K. was supported by the Shota Rustaveli National Science Foundation (Grant No. FR-23-856).

\end{acknowledgments}

\bibliography{/Users/phbb/Physics/Papers/refn} 

\end{document}